\documentclass[%
 reprint,
 amsmath,amssymb,
 aps,
 pre,
floatfix,
longbibliography
]{revtex4-1}
\usepackage{amsmath}
\usepackage{notes2bib}
\usepackage{soul}
\usepackage{graphicx}
\usepackage{dcolumn}
\usepackage{bm}
\usepackage[T1]{fontenc}
\usepackage[dvipsnames]{xcolor}
\newcommand{\authormain}{Farshid Jafarpour}
\newcommand{\titlemain}{Competition between transient oscillations and early stochasticity in exponentially growing populations}
\usepackage[bookmarks={false}, pdfauthor={\authormain}, pdftitle={\titlemain}]{hyperref}
\hypersetup{colorlinks=true, linkcolor=BrickRed, citecolor=Violet, filecolor=OliveGreen, urlcolor=RoyalBlue, filebordercolor={.8 .8 1}, urlbordercolor={.8 .8 0}}

\newcommand{\eq}[1]{Eq.~\eqref{eq:#1}}

\newcommand{\fig}[1]{Fig.~\ref{fig:#1}}
\newcommand{\Fig}[1]{Figure~\ref{fig:#1}}
\newcommand{\rf}[1]{Ref.~\cite{#1}}

\newcommand{\apx}[1]{Appendix~\ref{sec:#1}}

\begin{document}

\title{\titlemain}

\author{Ya\"ir Hein}
\affiliation{
	Institute for Theoretical Physics,
	Utrecht University, 3584 CC Utrecht, Netherlands}
\author{Farshid Jafarpour}%
\affiliation{
	Institute for Theoretical Physics,
	Utrecht University, 3584 CC Utrecht, Netherlands}

\date{\today}

\begin{abstract}
    It has been recently shown that the exponential growth rate of a population of bacterial cells starting from a single cell shows transient oscillations due to early synchronized bursts of division. These oscillations are enhanced by cell size regulation and contain information about single-cell growth statistics. Here, we report a phase transition in these oscillations as a function of growth rate variability. Below the transition point, these oscillations become asymptotically deterministic and can be measured experimentally, while above the transition point, the stochasticity in population growth dominates the oscillations and masks all the information about single cell growth statistics. The analytically calculated transition point, which roughly corresponds to $13\%$ variability in single-cell growth rate, falls within physiologically relevant parameters. Additionally, we show that the oscillations can stochastically emerge even when the initial state contains multiple cells with out-of-phase division cycles. We show that the amplitude and the phase of these oscillations are stochastic and would vary across repeated measurements with the same initial conditions. We provide analytic expressions as well as numerical estimates for the typical oscillation amplitude and the number of generations before the amplitude falls below a given measurement threshold for \textit{E. coli} in multiple growth conditions. 
\end{abstract}
\maketitle

\section{Introduction}

The relationship between bacterial populations and single-cell statistics has extensively been researched, particularly for cells in unrestricted growth~\cite{jafarpour2019cell,jafarpour2018bridging,hashimoto2016noise,levien2020interplay,lin2017effects,ho2018modeling,lin2020single,luo2021master,powell1956growth,nozoe2020cell,GAVAGNIN20211314,thomas2017single,chiorino2001desynchronization,tuanase2008regulatory,thomas2018analysis,garcia2019linking,levien2020large,vittadello2019mathematical,marantan2016stochastic,kendall1948role,hirsch1966decay,genthon2022analytical,barber2021modeling,jia2021cell,hein2022asymptotic}. In the presence of sufficient nutrients, bacterial populations grow exponentially at a constant rate that can be linked to the single-cell statistics \cite{hashimoto2016noise,levien2020interplay,lin2017effects,ho2018modeling,lin2020single,powell1956growth,tuanase2008regulatory,thomas2018analysis,thomas2017single,marantan2016stochastic,jia2021cell}. However, when the population starts from a few or a single cell, it takes some time before bacterial cell count grows perfectly exponentially, even if all individual cells are in balanced growth, meaning that their physiological states are adapted to the growth medium. 
This transient period is characterized by slowly decaying oscillations
~\cite{kendall1948role,nozoe2020cell,GAVAGNIN20211314,vittadello2019mathematical,chiorino2001desynchronization,yates2017multi,jafarpour2019cell,hein2022asymptotic}. These oscillations are distinct from the type of oscillations that arise due to cell interactions, such as resource competitions and stress signals \cite{martinez2018bistable}. Instead, these oscillations occur because the descendants of a cell grow together and divide more or less at the same time resulting in synchronized bursts of division in the population. As time passes, the stochasticity in cell growth slowly de-synchronizes the timing of cell division, which dampens the oscillations~\cite{jafarpour2019cell,hein2022asymptotic}.  
\\

To illustrate how transient behavior manifests, we focus on the cell count growth rate $\Lambda(t)$, defined as the fraction of cells in the population that divide per unit time $\frac{1}{N(t)}\frac{dN(t)}{dt}$. This population observable, along with many others, converges to a certain asymptotic constant value after an infinite amount of time. At large but finite times, however, these population observables exhibit transient behavior. The ensemble averages of intensive observables oscillate around their asymptotic value, with an exponentially decaying amplitude. The time scale for this oscillation decay has been derived for several single-cell models of growth and division, and it scales with the variability of single-cell growth rates \cite{kendall1948role,nozoe2020cell,GAVAGNIN20211314,vittadello2019mathematical,chiorino2001desynchronization,yates2017multi,jafarpour2019cell,hein2022asymptotic}. On the other hand, due to stochasticity in single-cell growth and division, cell count $N(t)$ evolves stochastically with fluctuations that scale with $\sqrt{N(t)}$. The relative amplitude of such fluctuations, therefore, scales as $1/\sqrt{N(t)}$, which decays exponentially at a rate of half the population growth rate. In other words, both the oscillations and fluctuations in cell count growth rate decay exponentially with competing decay rates.
\\

There is a transition point at which the transient oscillations and fluctuations are equal. Since the oscillations and fluctuations decrease in magnitude at different rates, the transient population behavior will ultimately be dominated by only one of these effects: the one with the lowest decay rate. We can, therefore, formulate a condition necessary for the oscillations to dominate the transient population behavior. In \rf{jafarpour2019cell}, the oscillation decay rate was estimated to be $r\approx 29\, \sigma_\kappa^2/\bar\kappa$, where $\sigma_\kappa$ and $\bar\kappa$ are the single-cell growth rate standard deviation and mean respectively. The asymptotic population growth rate is roughly identical to the single-cell growth rate $\bar\kappa$, so the fluctuations decay at a rate of $\bar\kappa/2$. We can therefore say that oscillations dominate the transient behavior only if 
\begin{equation}
\label{eq:crossover_0}
    \sigma_\kappa \leq 0.13\, \bar\kappa.
\end{equation}
This transition point lies within the range of physiological values for $\sigma_\kappa$ and $\bar\kappa$ \cite{taheri2015cell}.
\\

The aforementioned transition point was defined for a model in which growth rate correlations were neglected. Population behavior depends on how cell division times are distributed and how they are correlated between mother and daughter cells \cite{lin2017effects,lin2020single}. A good way to model these correlations is by treating growth rate variability and division size variability as two separate sources of noise, each with its own mother-daughter cell correlations \cite{taheri2015cell}. In \rf{hein2022asymptotic} the authors use such a description of single-cell behavior to quantitatively predict the oscillatory population behavior. They assume that cell growth rate along a lineage is a continuous random process. The per-cell growth rate variability and correlations then emerge from the noise and auto-correlation of the underlying growth process. All of their results have, however, only been derived for ensemble averaged populations, which means they only considered observables corresponding to averages over repeated experiments with the same initial conditions.
\\

In this paper, we adopt the model from \rf{hein2022asymptotic},  but consider the behavior of single population experiments instead of just the ensemble average, which causes some interesting effects that are often ignored in other literature. Unlike in the ensemble average population, we observe a transition between fluctuation-dominated and oscillation-dominated transient regimes. We determine a generalized transition point, of which \eq{crossover_0} is a special case. Another interesting phenomenon absent in the ensemble average is that the amplitude of the transient oscillations is random among repeated experiments, set by stochasticity in the early population. We analyze the distribution of these transient oscillation amplitudes through simulations and theory. Interestingly, the typical amplitude of a single population is higher than that of an ensemble-averaged population, due to the net effects of spontaneous cell synchronization. Our predictions are in great qualitative agreement with oscillations studied in a multistage model used to describe cancer cell populations \cite{GAVAGNIN20211314}. At last, we consider tables of realistic parameters based on data from \rf{taheri2015cell} and answer just how many generations it typically takes for the oscillations to fall below a certain threshold.

\section{Model}
Suppose we have a population that starts from a single cell. We assume this cell has already been adapted to the growth medium. We want to keep track of the number of cells $N(t)$. For this, we use a physiologically relevant model of growth and division, which accounts for growth fluctuations, their correlations, and cell size regulation based on the model from \rf{hein2022asymptotic}. The effect of asymmetric division is ignored, since its effect of oscillations is relatively small for physiologically relevant values \cite{jafarpour2019cell,eun2018archaeal,campos2014constant,guberman2008psicic}.
\\

Consider the following model. At each point in time, there are a total of $N(t)$ cells in the population, each of which can be labeled by an index $j \in \{1,\dots, N(t)\}$. Every cell $j$ can be characterized by two variables, its cell mass $m_j(t)$ (or equivalently cell size, volume, area, etc.) and growth state $\mathbf x_j(t)$. The cell mass grows exponentially in time, meaning it satisfies
\begin{equation}
\label{eq:ddtvjt}
    \frac{d}{dt} m_j(t) = \lambda(\mathbf x_j(t))m_j(t),
\end{equation}
where the growth rate $\lambda(\mathbf x)$ is a function of the growth state $\mathbf x$, which is an abstract vector that represents the internal state of the cell. The growth state also changes stochastically in time, with each cell's growth state fluctuating independently of the other cell's. For a single cell, we denote the probability of finding a cell in growth state $\mathbf x$ at time $t$ is given by $p(t,\mathbf x)$. This probability evolves according to
\begin{equation}
\label{eq:ddtp}
    \frac{\partial}{\partial t} p(t,\mathbf x) = \mathcal{K} p(t,\mathbf x),
\end{equation}
where $\mathcal{K}$ is some differential operator. We also assume that at any moment cells have a chance of dividing. The rate of division may be any function of cell mass $m$ and the masses at previous births and divisions, as long as the distribution of cell mass at division goes to a steady state within a reasonable amount of time. Upon division, two new daughter cells are formed with the same growth state but half the mass of the mother cell at division. 
\\

When concrete examples are needed, we will assume $\lambda_t= \lambda(\mathbf x(t))$ to be an Ornstein-Uhlenbeck process, which is characterized by three parameters. These are the mean $\bar\lambda = \langle \lambda_t\rangle$, variance $\sigma_\lambda^2 = \text{Var}(\lambda_t)$ and correlation time $\tau_{cor}$ which sets the correlation decay time via $\text{Cov}(\lambda_t,\lambda_s)/\sigma_\lambda^2 = e^{-|t-s|/\tau_{cor}}$. There is a direct correspondence between the average division time along a lineage and the mean growth rate set by $\tau_{div} = \ln(2)/\bar\lambda$. 
\\


When concrete models of division are needed, we assume that cell masses at division $m_d$ depend on their masses at birth $m_b$, via the relation based on the cell size regulation model from \rf{amir2014cell},
\begin{equation}
\label{eq:lnmd}
    \ln(m_d) = \ln(2 \bar m) + (1-\alpha)\ln(m_b/\bar m) + \eta,
\end{equation}
where $\alpha$ is the mother-daughter division mass correlation parameter, $\eta$ is a normally distributed random variable with mean zero and variance $\sigma_\eta^2$, and $\bar m$ is a constant cell mass that corresponds to an equilibrium birth mass. We assume symmetric division, so the daughter cells' birth masses $m_d$ are set to be half their mother cell's mass at the division. After a few generations, the mass at birth and division quickly settle to equilibrium log-normal distributions with $\langle \ln m_b \rangle = \ln \bar m$ and $\langle \ln m_d\rangle = \ln 2\bar m$ respectively~\cite{hein2022asymptotic}. This motivates calling $\bar m$ the equilibrium birth mass. The coefficients of variation of birth mass and division mass are identical and given by
\begin{equation}
    \text{CV}_{m_b} \approx \sqrt{\text{Var}(\ln m_b)} = \frac{\sigma_\eta}{\sqrt{\alpha(2-\alpha)}}.
\end{equation}
In our simulations, we always use $\alpha=0.5$, which corresponds to an adder model of cell size regulation~\cite{amir2014cell,taheri2015cell}. The simulation code is available at \cite{Code}\\

We emphasize that our analytical results do not assume this particular choice of cell size regulation model used in the simulation. We will show that the only aspect of the cell division model that affects long-term population dynamics is the value of $\text{CV}_{m_d}$.

\section{Ensemble averaged oscillations and Oscillation-fluctuation trade-off}
\label{sec:Ens}
If one could repeat the same experiment with the same initial conditions and take the average over all cell count trajectories, one would obtain some ensemble average $\langle N(t)\rangle$. For this function, we define the ensemble population growth rate
\begin{equation}
\label{eq:LN}
    \Lambda_{\text{ens}}(t) = \frac{1}{\langle N(t)\rangle} \frac{\text{d}\langle N(t)\rangle}{\text{d} t},
\end{equation}
This effective growth rate is known to converge to an asymptotic population growth rate $\Lambda_\infty$ at large times~\cite{powell1956growth,lin2017effects,lin2020single}. This limit holds true when the cell mass and growth state distribution of the population have reached a steady state, in which case the cell masses have completely de-synchronized. Alternatively, we could define the population growth rate as the change in the average logarithm of population count $d\left\langle\ln N(t)\right\rangle/dt$ \cite{lewontin1969population}. We use \eq{LN} because it is easier to work with analytically while it still exhibits the same oscillatory transient behavior of interest.
\\

Let us explain why a non-steady-state population exhibits oscillations in population growth rate. When the average cell mass exceeds the steady-state average, we have an increase in cells close to division, which momentarily raises $\Lambda(t)$ with respect to the steady-state value $\Lambda_\infty$. Following an increase in divisions, the abundance of newborn cells causes average cell mass to drop below the steady state, thereby decreasing $\Lambda_{\text{ens}}(t)$. When the number of divisions $\Lambda_{\text{ens}}(t)$ is low, average cell mass builds up again and the cycle repeats. This process causes wave-like behavior in cell mass distribution and cell division rate $\Lambda_{\text{ens}}(t)$.  The average effect of cells growing at different rates de-synchronizes the population. This dampens the mass synchronicity waves, at rates proportional to their oscillation speed. This effect rapidly dampens out any higher-order mass synchronicity oscillations. What is left are just the first-order oscillations corresponding to full cell cycles. The transient regime is thus well described by just the first-order oscillations ~\cite{hein2022asymptotic}
\begin{equation}
\label{eq:LEN}
    \Lambda_{\text{ens}}(t) \approx \Lambda_\infty \left(1 + A_{\text{ens}} e^{-r t}\cos(\Omega t + \phi_{\text{ens}})\right),
\end{equation}
Here, the amplitude pre-factor $A_{\text{ens}}$ and phase $\phi_{\text{ens}}$ are constants that depend on the division process, growth process as well as the population's initial conditions. An explicit expression for $A_{\text{ens}}$ is given in Section \ref{sec:typical_amplitude}. The asymptotic growth rate $\Lambda_\infty$, oscillation decay rate $r$, and oscillation speed $\Omega$, however, are time scales that are uniquely determined by the model of cell growth. In \rf{hein2022asymptotic} these timescales were given in the case where $\lambda_t$ is an Ornstein-Uhlenbeck process. Here, we provide a more general expression. 
\\

Assuming a growth rate process $\lambda(\mathbf x)$ where $\mathbf x$ evolves according to \eq{ddtp}, we find that $\Lambda_\infty$ is the leading eigenvalue of the operator $\mathcal K + \lambda(\mathbf x)$. Let $\mu$ be the leading eigenvalue by real part of the operator $\mathcal  K + [1+i 2\pi/\ln(2)] \lambda(\mathbf x)$. Then we have that the oscillation decay rate is $r= \Lambda_\infty -\text{Re}\mu$ and the oscillation speed is $\Omega = \text{Im} \mu$. These expressions reduce to forms that are easier to interpret if we assume $\lambda_t$ is Ornstein-Uhlenbeck or assume small growth rate variability $\sigma_\lambda \ll \bar\lambda$. For any growth rate process, one can define an auto-correlation function
\begin{equation}
    \rho(s):= \text{Cov}(\lambda_t,\lambda_{t+s})/\sigma_\lambda^2.
\end{equation}
A key variable that links the population time-scales to the growth process is the integrated correlation
\begin{equation}
\label{eq:D}
    D:=  \sigma_\lambda^2 \int_0^\infty ds\rho(s).
\end{equation}
This parameter acts as the effective diffusion constant of the integrated growth rate process over time scales much larger than the correlation time $\tau_{\text{cor}}$. For a growth process $\lambda_t$ that is Ornstein-Uhlenbeck or has small growth rate variability $\sigma_\lambda \ll \bar\lambda$, the population time scales are 
\begin{itemize}
    \item population growth rate $\Lambda_\infty = \bar\lambda + D$
    \item oscillation decay rate $r = \frac{4\pi^2}{\ln(2)^2} D \approx 82D$
    \item oscillation speed $\Omega = \frac{2\pi}{\ln(2)}(\bar\lambda+2D)$.
\end{itemize}
When $\lambda_t$ is Ornstein-Uhlenbeck, we have that $D=\sigma_\lambda^2 \tau_{\text{cor}}$. A derivation of the time scales is shown in Appendix~\ref{sec:small}. When considering collective oscillations, small phase shifts among the oscillations of sub-populations can greatly affect the amplitude of their sum due to destructive interference. This is reflected in the large constant $4\pi^2/\ln(2)^2\approx 82$ that connects the oscillation amplitude decay rate $r$ to growth fluctuations. Any large constants encountered in the rest of the text result from similar effects and their exact form can be found in the appendix.

\begin{figure*}
\centering
\includegraphics[scale=0.45]{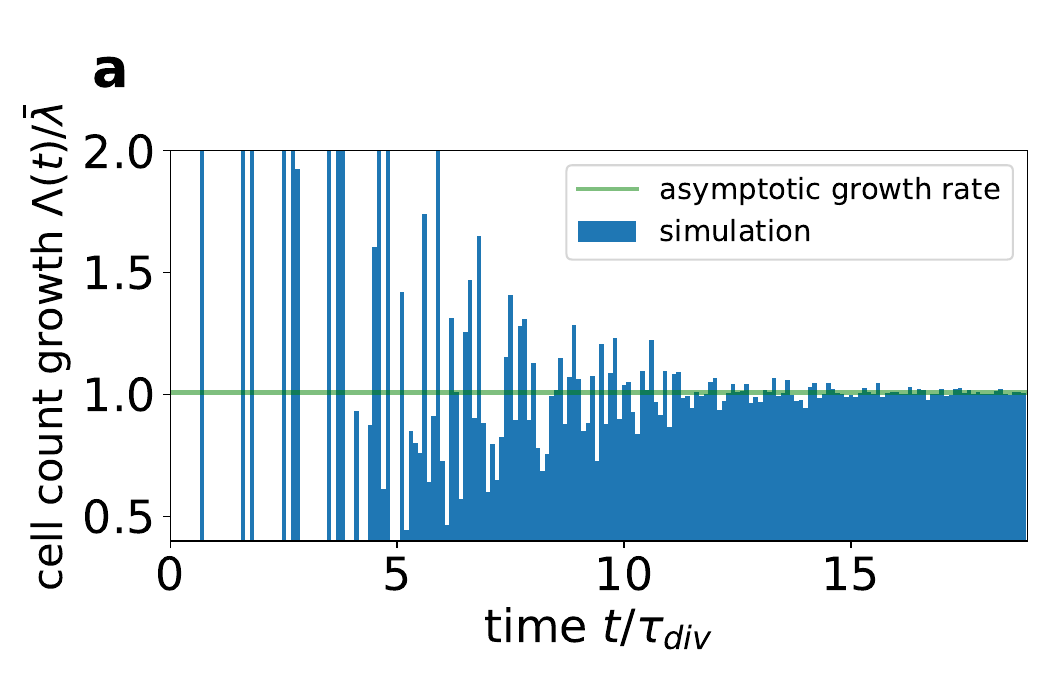}
\includegraphics[scale=0.45]{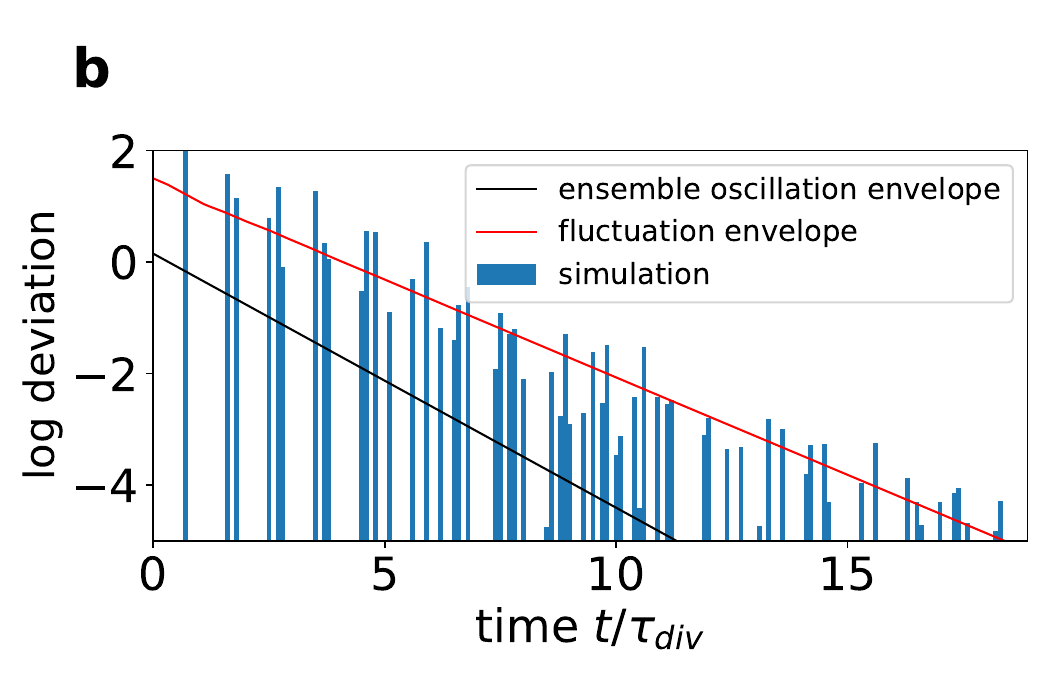}
\includegraphics[scale=0.45]{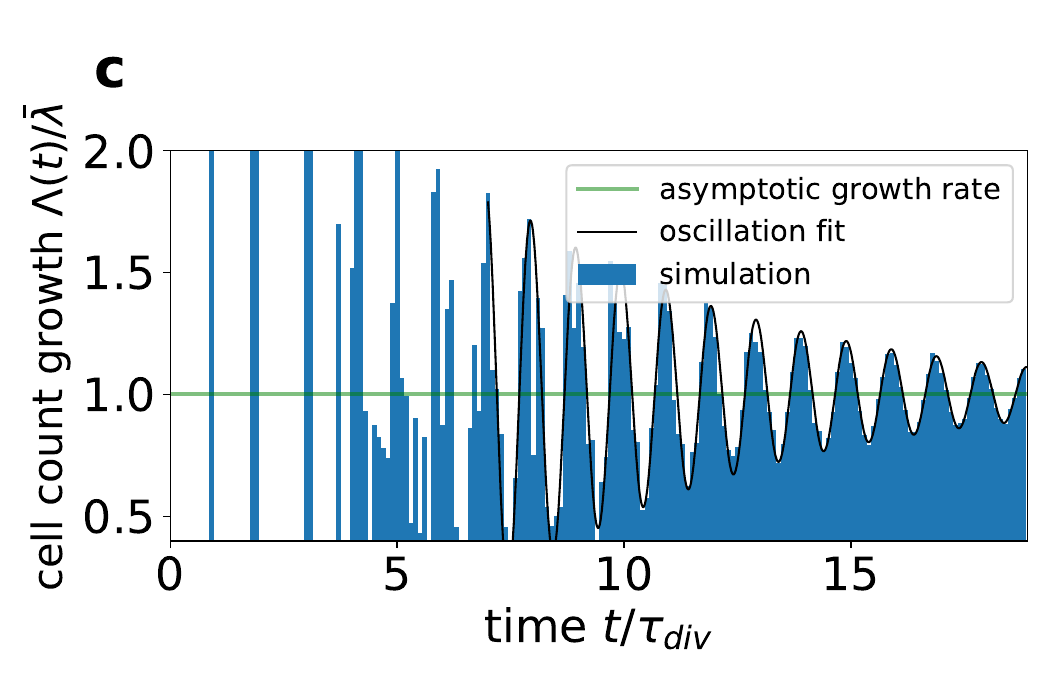}
\includegraphics[scale=0.45]{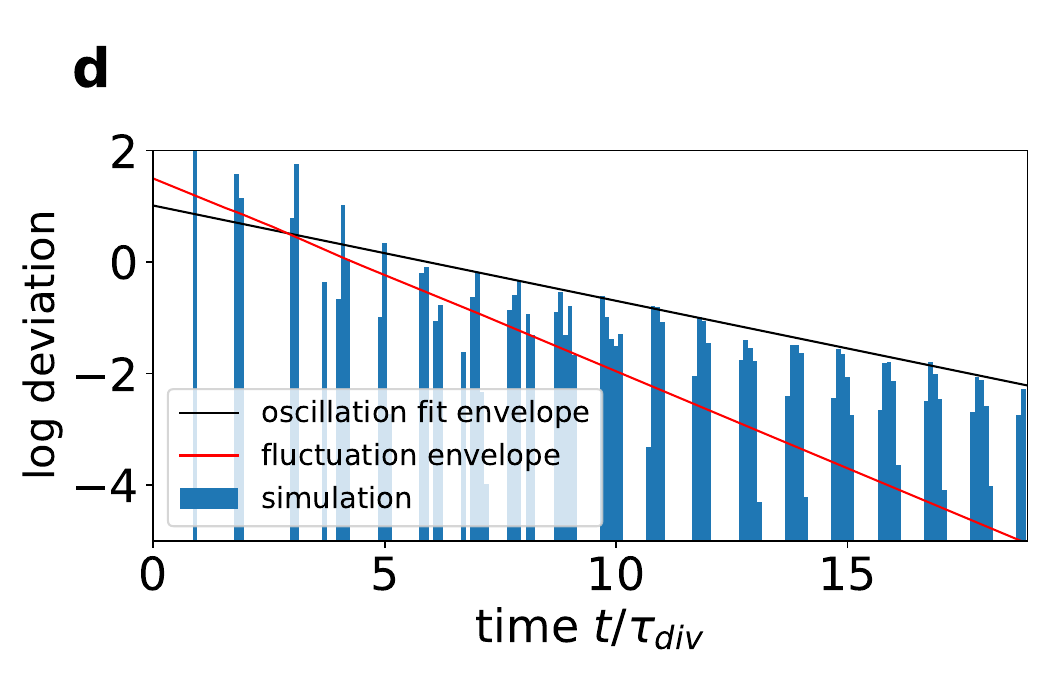}
\caption{Population growth rate as a function of time for two parameter values on both sides of the transition. (a) growth rate above the transition is stochastic and approaches steady exponential growth, while (c) the growth rate below the transition approaches a decaying oscillation around the exponential growth. (b) and (d) are the same plots in logscale (with asymptotic values subtracted, $\ln_{>0}[(\Lambda(t)-\Lambda_\infty)/\bar\lambda]$) showing the exponential decay of transient dynamics. In (b), above the transition, oscillations decay faster than the fluctuations, while in (d) the oscillations decay slower than the fluctuations and become asymptotically deterministic. Simulation parameters: $\sigma_\lambda^2 \tau_{\text{cor}}/\bar\lambda =0.008$, which is $\epsilon=1.3$ for (a) and (b), and $\sigma_\lambda^2 \tau_{\text{cor}}/\bar\lambda =0.003$, which is $\epsilon=0.5$ for (c) and (d). All simulations start from a single cell with cell mass $m_0=\bar m=1$ and growth rate $\lambda_0 =\bar\lambda =\ln(2)$ (such that $\tau_{\text{div}}=1$). Simulation details: Given a cell's birth size, its target division size is randomized according to \eq{lnmd}. with division parameters $\text{CV}_{m_d}=0.12$ and $\alpha=0.5$. At every time increment of $dt=0.01$, each cell's growth rate growth rate is updated as an Ornstein-Uhlenbeck process, $\lambda_{t+dt} = e^{-dt/\tau_{\text{cor}}}\lambda_t + \sqrt{2\sigma_\lambda^2/\tau_{\text{cor}}} dW_t$, where $dW_t$ is drawn from a normal distribution such that $\mathbb E [dW_t] = 0$ and $\text{Var}(dW_t)=dt$. Cell masses are updated as $m_{t+dt} = m_t e^{\lambda_t dt}$. After every time step, for every cell whose mass now exceeds the division mass $m_t >m_d$, we take that cell out of the population and replace it by two daughter cells with birth size $m_t/2$ and the same growth rate $\lambda_t$ as their mother cell. Throughout the simulation, we track the number of cells $N(t)$ at every time step. Population growth rate $\Lambda_t$ is determined using \eq{L_Delta_def} with a bin size of $\Delta t = 0.1$. In Appendices \ref{sec:fluctuation envelope} and \ref{sec:osc_fit} we discuss how we obtained the fluctuation and oscillation fit envelopes respectively. The simulation code is provided in the GitHub repository~\cite{Code}. }
\label{fig:transition}
\end{figure*}

\subsection*{Oscillation-fluctuation transition}
In a single population experiment/simulation, depending on the parameter regime, the transient behavior is dominated either by fluctuations or deterministic oscillations of the form in \eq{LEN}. For a population with cell count trajectory $N(t)$, we define the cell count growth rate as
\begin{equation}
\label{eq:Lt_def}
    \Lambda(t) := \frac{1}{N(t)} \frac{\text{d}N(t)}{\text{d}t}.
\end{equation}

Suppose that the $N(t)$ cells in a population grow independently. The fluctuations in the number of cells that divide per unit time $dN(t)/dt$ now scale with $\sqrt{N(t)}$. Consequently, the fluctuations in the cell count growth rate $\Lambda(t)$ scale with $1/\sqrt{N(t)}$, which evolves as $e^{-\frac{\Lambda_\infty}{2} t}$. This introduces a timescale of $\Lambda_\infty/2$ in addition to the oscillation decay rate $r$. We define the decay ratio $\epsilon$ as the ratio of the oscillation decay rate to the fluctuation decay rate
\begin{equation}
\label{eq:eps}
    \epsilon := \frac{r}{\Lambda_\infty/2} \approx 164 \sigma_\lambda^2 \tau_{\text{cor}}/\bar\lambda.
\end{equation}
If $\epsilon <1$, then the oscillations decay more slowly than the background fluctuations. In this regime, the ratio of fluctuations to oscillations in $\Lambda(t)$ will always go to zero eventually. When this happens, the dynamics of $\Lambda(t)$ become deterministic. Conversely, if $\epsilon \geq 1$, then all oscillations will eventually fall below the background fluctuation level. the trajectory of $\Lambda(t)$ is dominated by noise and local oscillations caused by spontaneous cell mass synchronicity. In \apx{crossover_0} we show how in the limit of vanishing growth rate correlations, the condition $\epsilon <1$ is equivalent to \eq{crossover_0}.
\\

In \fig{transition} we show examples of the trajectories of $\Lambda(t)$ for $\epsilon=0.5$ and $\epsilon=1.3$, as well as the log of the deviation of $\Lambda(t)$ from its asymptotic value $\Lambda_\infty$. The oscillation fit and envelope are obtained by taking the theoretical transient oscillatory behavior and matching the amplitude and phase to the cell mass distribution at the end of the simulation. The fluctuation envelope scales with $\sqrt{N(t)}$ (see Appx. \ref{sec:envelopes} for precise definitions and derivations of the oscillation and fluctuation envelopes).
In log deviation space, the exponential decay rates of transient effects appear as slopes. Which transient effect dominates in the large time limit only depends on the theoretical ratio of the slopes set by $\epsilon$. One can think of the point at which the oscillations clearly separate from the background noise as the crossover between the two lines. At the transition, as $\epsilon$ goes to $1$, this separation time diverges. In Sec. \ref{sec:analogy} we will show that this dynamic transition at $\epsilon=1$ has properties that are analogous to equilibrium phase transitions.
\\

\section{Oscillation amplitude and phase vary across repeated identical experiments}
\begin{figure}
    \includegraphics[scale=0.5]{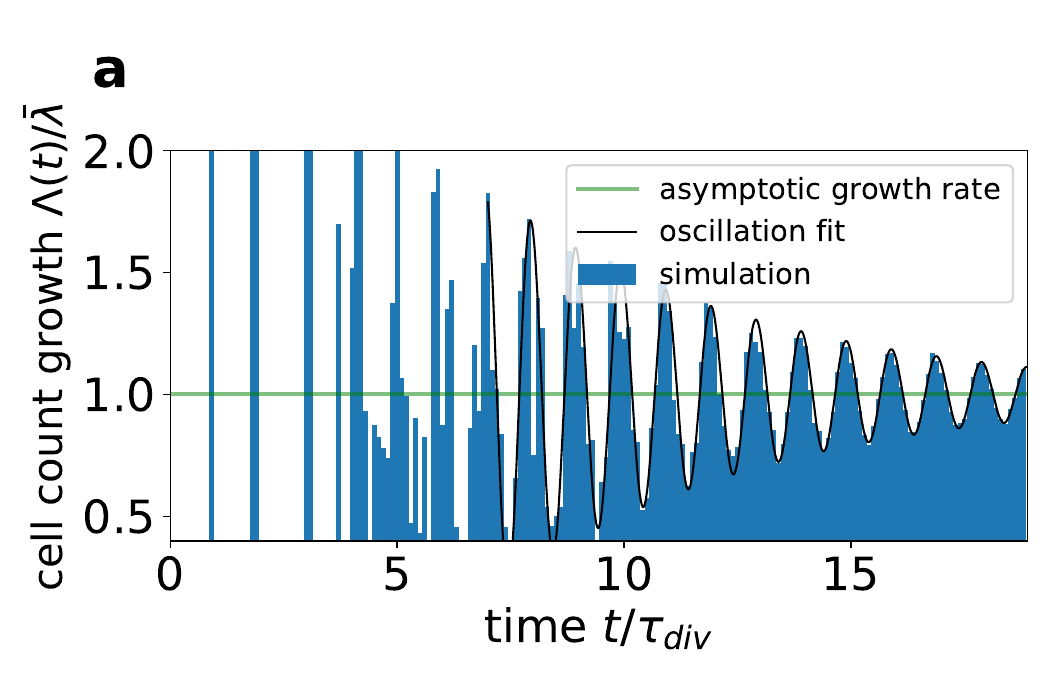}
    \includegraphics[scale=0.5]{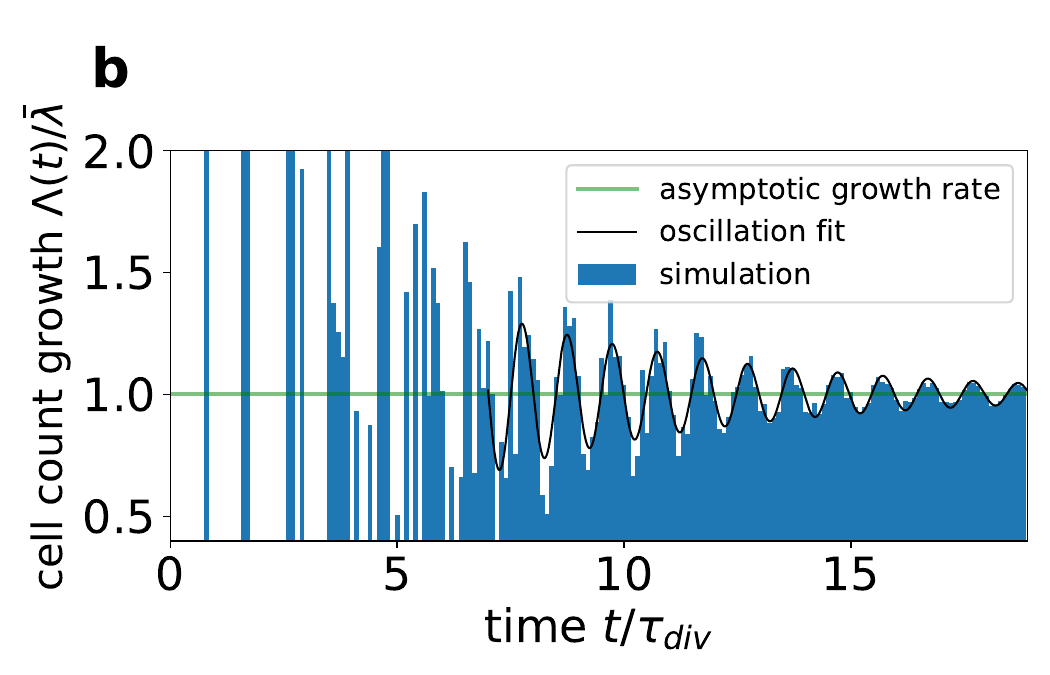}
    \caption{(a) and (b) show two trajectories of cell count growth rates $\Lambda(t)/\bar\lambda$ 
    for two simulations with identical parameters and initial conditions. Note how the oscillations have different amplitudes despite both simulations having identical initial conditions. Simulation details: starting with one cell with mass $m(0)=\bar m =1$ and growth rate $\lambda_0=\bar\lambda= \ln(2)$ with growth parameters $\tau_{\text{cor}}=5/\bar\lambda$, $\sigma_\lambda^2 \tau_{\text{cor}}/\bar\lambda =0.003$ and division parameters $\text{CV}_{m_d}=0.12$, $\alpha=0.5$. This corresponds to a decay ratio of $\epsilon = 0.5$.}
    \label{fig:osc_fits}
\end{figure}

In this section, we consider the case of $\epsilon<1$ and discuss the effects of fluctuations on the oscillations in single-population experiments. \Fig{osc_fits} (a) and (b) show the trajectories of $\Lambda(t)$ for two populations with the same parameters and initial conditions. In all simulations, we observe transient oscillations of the form
\begin{equation}
\label{eq:Lt}
    \Lambda(t) \approx \Lambda_\infty\left( 1+ A e^{-rt}\cos(\Omega t + \phi)\right).
\end{equation}
Although time scales $\Lambda_\infty$, $r$, and $\Omega$ are the same as for the ensemble's oscillations, the amplitude $A$ and phase $\phi$ are now random variables that are set by the population's early history, which is in close agreement with observations from \rf{GAVAGNIN20211314}. In \fig{A_dist} (a) and (b), we show the spread in the instantaneous amplitude (see \apx{osc_amp} for the definition) after 12 generations. Interestingly, the ensemble average amplitude $A_{\text{ens}}$ from \rf{hein2022asymptotic} visibly underestimates the typical amplitude of a single population experiment. 

\subsection*{The typical amplitude is larger than the amplitude of the ensemble average}
\label{sec:typical_amplitude}
In this section, we de-generalize our model a little to investigate how the amplitude prefactor $A$ in \eq{Lt} is distributed and depends on the model's parameters. We assume that the growth process $\lambda_t$ is Ornstein-Uhlenbeck with mean $\bar\lambda$, variance $\sigma_\lambda^2$, correlation time $\tau_{corr}$, and that the initial growth rate $\lambda_0$ is close to $\bar\lambda$. We also assume that the cell size at division admits some steady-state distribution with a coefficient of variation of $\text{CV}_{m_b}$. Based on the theory from \rf{hein2022asymptotic}, the ensemble average amplitude from \eq{LEN} of a population starting from a single cell is approximately given by
\begin{equation}
\label{eq:A_ens}
A_{\text{ens}} =2  e^{-41 \text{CV}_{m_b}^2+123 \sigma_\lambda^2 \tau_{cor}^2}.
\end{equation}
To summarize \eq{A_ens}, two major factors affect the ensemble average amplitude. Firstly, growth rate correlations increase this amplitude, as they cause some delay in the time it takes for cells to fluctuate independently and de-synchronize their masses. Secondly, variability in division mass decreases the amplitude, as this causes a local spread in cell division timings among cells with synchronized mass, thereby slightly flattening oscillations in $\Lambda(t)$. In \fig{transition2} (a), we plot $A_{\text{ens}}$ for various values of $\tau_{\text{cor}}$ and $\epsilon$ (using \eq{eps} to determine $\sigma_\lambda^2$) and no cell division variability $\text{CV}_{m_b}=0$. We see the dependence of $A_{\text{ens}}$ on the growth rate parameters is fairly limited in the relevant regime. 
\\

\begin{figure*}
    \centering
    \includegraphics[width=0.49\textwidth]{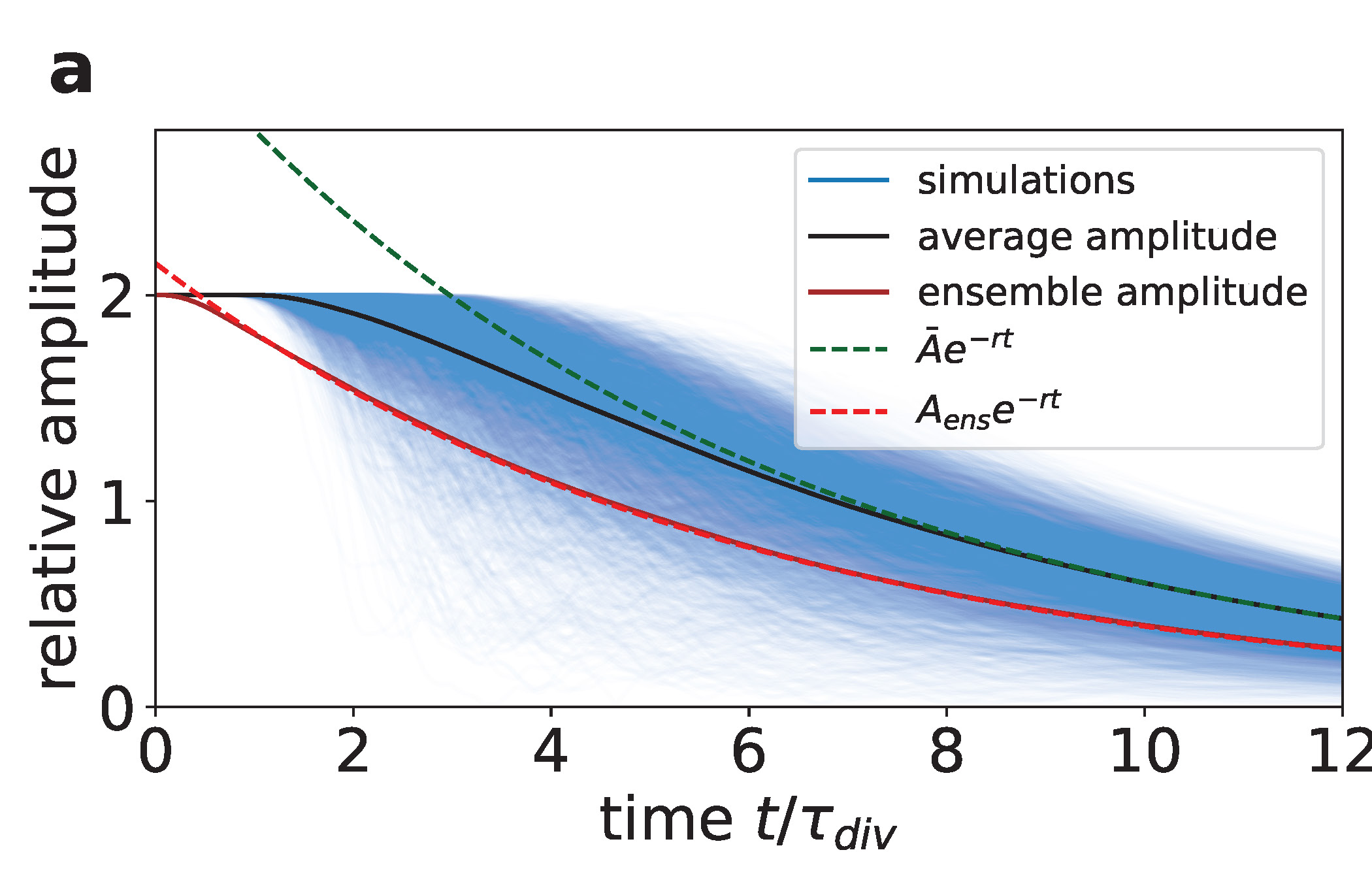}\hfill%
    \includegraphics[width=0.49\textwidth]{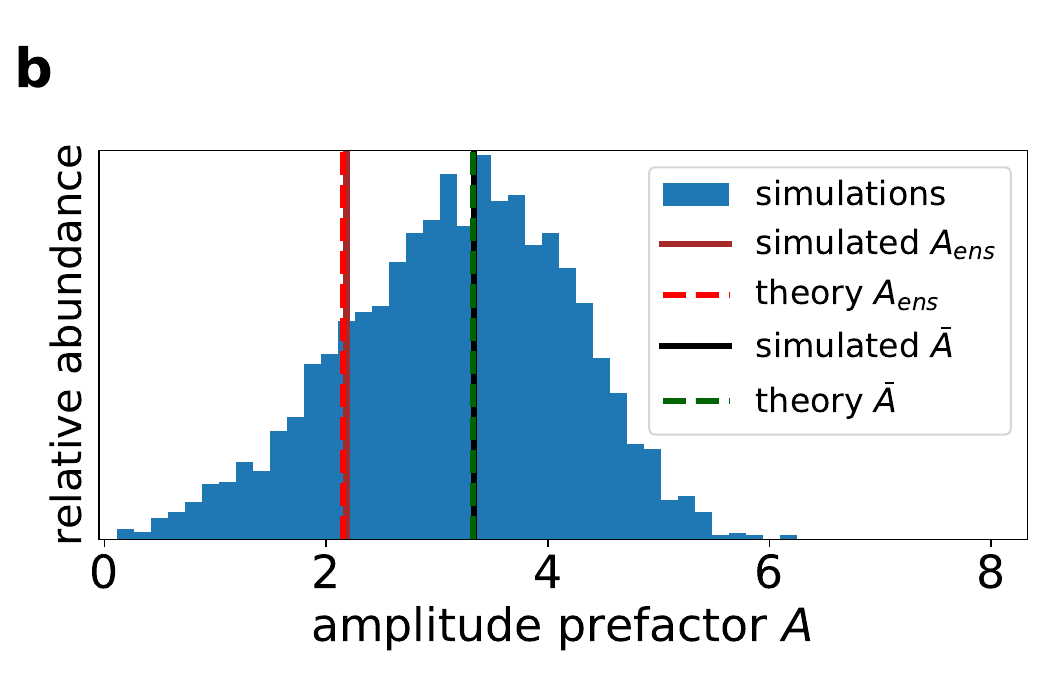}
    \includegraphics[width=0.49\textwidth]{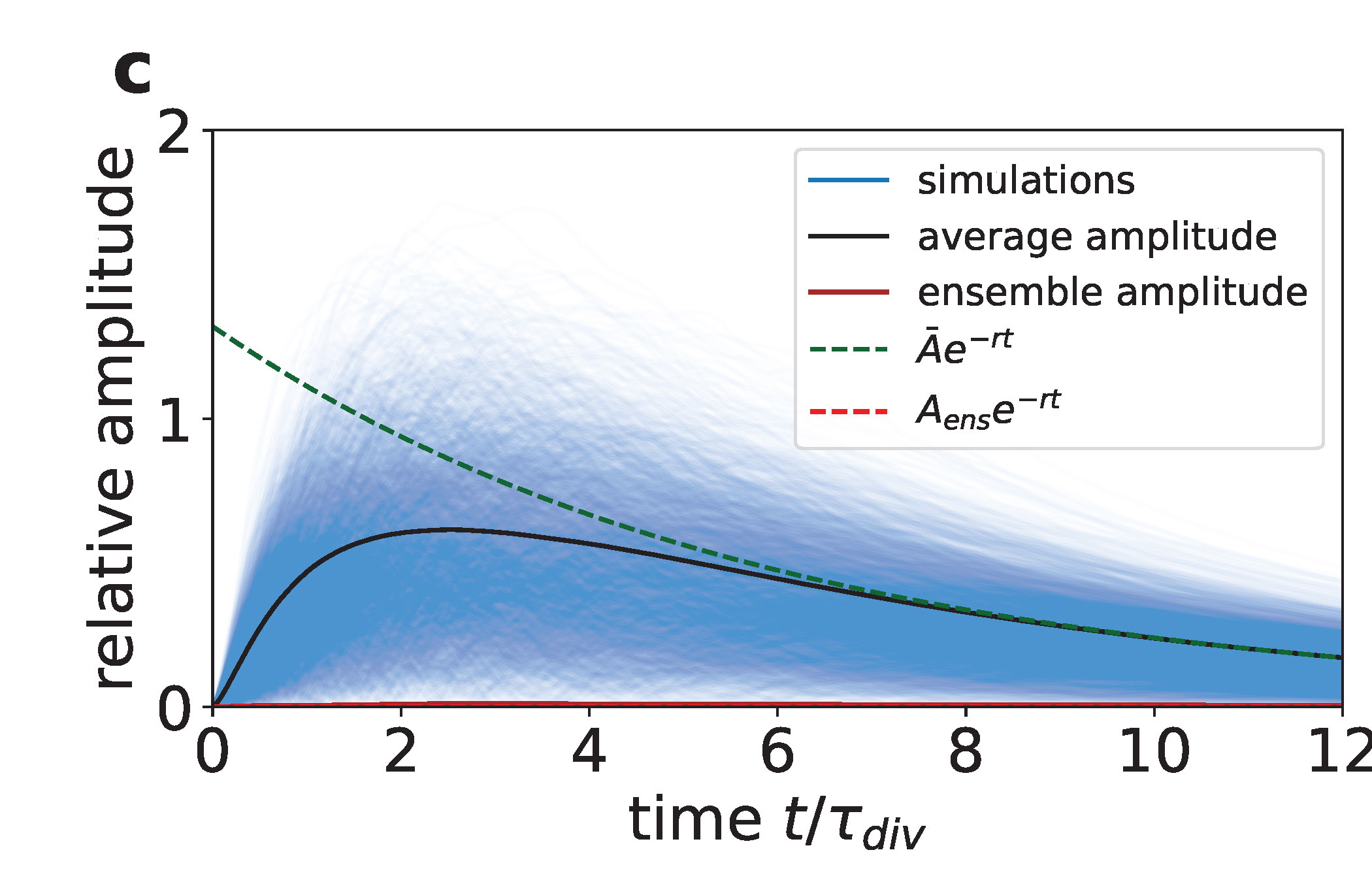}\hfill%
    \includegraphics[width=0.49\textwidth]{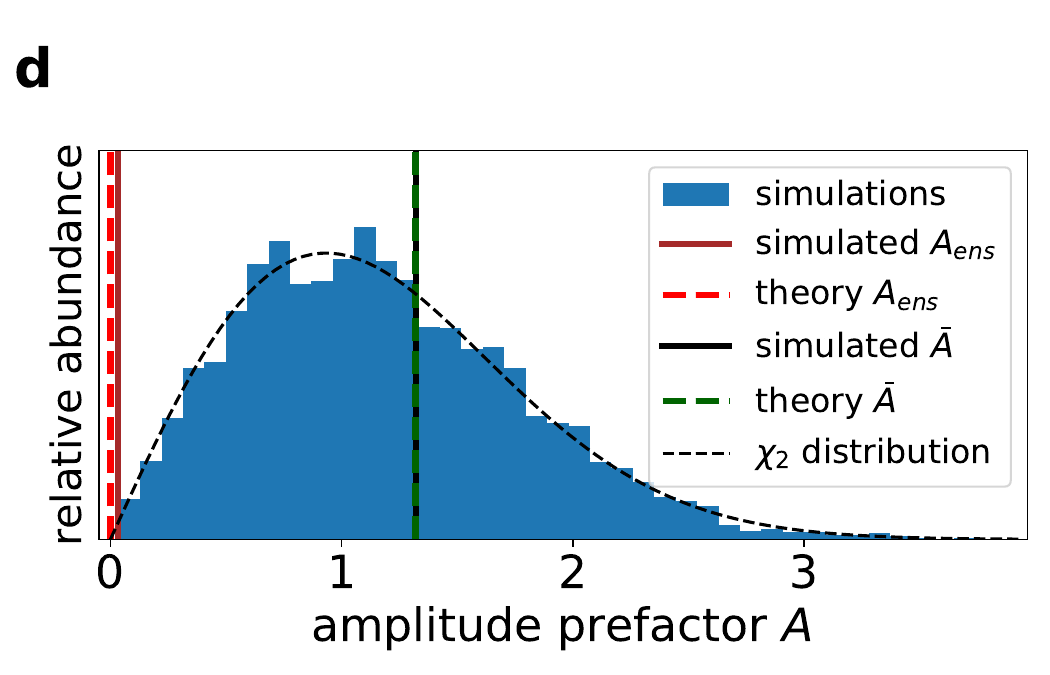}
    \caption{Instantaneous relative oscillation amplitudes in $\Lambda(t)$ for $4000$ simulations and their distribution. (a) and (b) represent the same set of simulations that started with a single cell, while in (c) and (d), the population started with three cells. The instantaneous amplitude is given by $A_{\text{inst}}(t) =e^{-41\text{CV}_{m_d}^2}|\Psi(t)|$, with $\Psi(t)$ given in \eq{Psi_def}. In (a) and (c), the black line is the root mean square $\sqrt{\langle A_{\text{inst}}(t)^2\rangle}$, and the green line is its asymptotic prediction $\bar A e^{-r t}$, where $\bar A$ is given by \eq{bar_A}. The brown solid line is the ensemble average's amplitude, and the red dashed line is its asymptotic prediction $A_{\text{ens}}e^{-rt}$, where $A_{\text{ens}}$ is given by \eq{A_ens}. In (b) and (d) we show the distribution of the amplitude pre-factor $A_{\text{inst}}(t_{max})e^{r t_{max}}$ based on their values at $t_{max}=12\tau_{\text{div}}$. Simulations in (c) and (d) start with three cells with masses $m(0)/\bar m = 1, 1.19,1.56$, chosen such that initial cell mass synchronicity is $\Psi(0)=0$. The remaining simulation details are identical to the simulations used to generate \fig{transition} (c) (d) and \fig{osc_fits}. }
    \label{fig:A_dist}
\end{figure*}

The transient oscillation amplitude differs per population, even for populations with the same initial conditions. For any population with sustained oscillations, there will be some random prefactor $A$ that describes the transient oscillations as in \eq{Lt}. To investigate the dependency of this amplitude on initial conditions and parameters, we consider a 'typical' amplitude pre-factor $\bar A$, defined as the root mean square
\begin{equation}
\label{eq:A_bar}
    \bar A := \sqrt{\mathbb E\left[ A^2\right]},
\end{equation}
where the average is taken over independent stochastic outcomes of the same initial population. Note the difference with $A_{\text{ens}}$ given in \eq{LEN}. The typical amplitude pre-factor $\bar A$ is obtained from averaging amplitudes of different realizations of $\Lambda(t) =  \dot N(t)/N(t)$, whereas the ensemble amplitude pre-factor $A_\text{ens}$ belongs to the evolution of $\Lambda_{\text{ens}}(t) = \langle \dot N(t)\rangle/\langle N(t)\rangle$.\\

In \apx{osc_amp} we show that this typical amplitude is closely approximated by
\begin{equation}
\label{eq:bar_A}
    \bar A = 2 e^{-41 CV_{m_b}^2 + 123\sigma_\lambda^2 \tau_{cor}^2} \sqrt{\frac{m_0^{-\epsilon}}{2^{1-\epsilon}-1}}.
\end{equation}
In \fig{A_dist} (a) and (b) we see how the typical amplitude pre-factor $\bar A$ more closely resembles the amplitude in typical experiments than $A_{\text{ens}}$. The right-most factor in \eq{bar_A} is strictly greater than one, and it quantifies the net increase in typical amplitude pre-factor with respect to the ensemble average due to stochasticity in the early population. In \fig{transition2} (b), we plot $\bar A$ for various values of $\epsilon$ and $\tau_{\text{cor}}$, showing that $\bar A$ diverges at the oscillation-fluctuation phase transition as $\epsilon$ goes to $1$. Note that this does not mean that we observe arbitrarily large oscillations, since an increase in $\epsilon$ also means that oscillations decay faster. The instantaneous oscillation amplitude at a fixed point in time therefore still goes to a constant as $\epsilon$ goes to $1$.

\subsection*{Analogy with equilibrium phase transition}\label{sec:analogy}
\begin{figure*}
\centering
\includegraphics[width=0.32\textwidth]{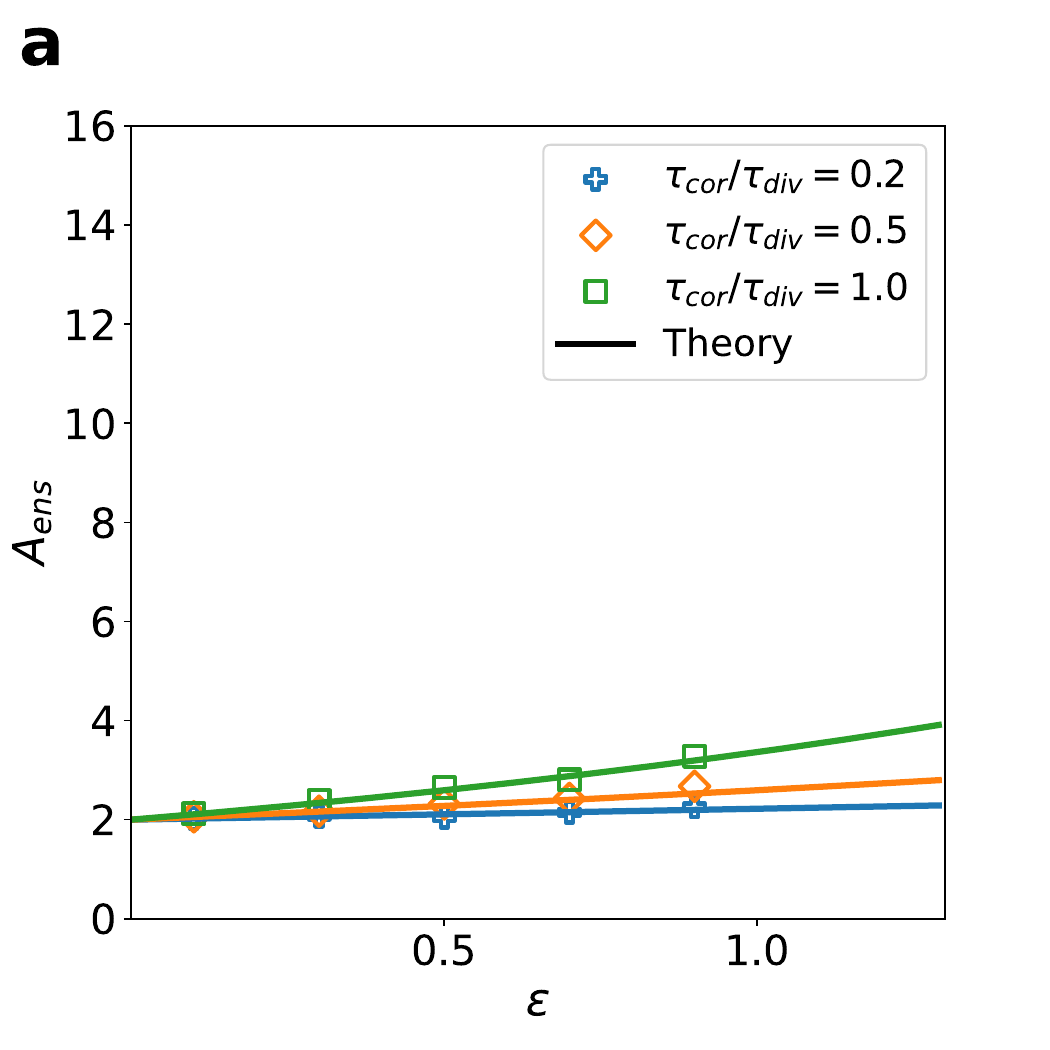}
\includegraphics[width=0.32\textwidth]{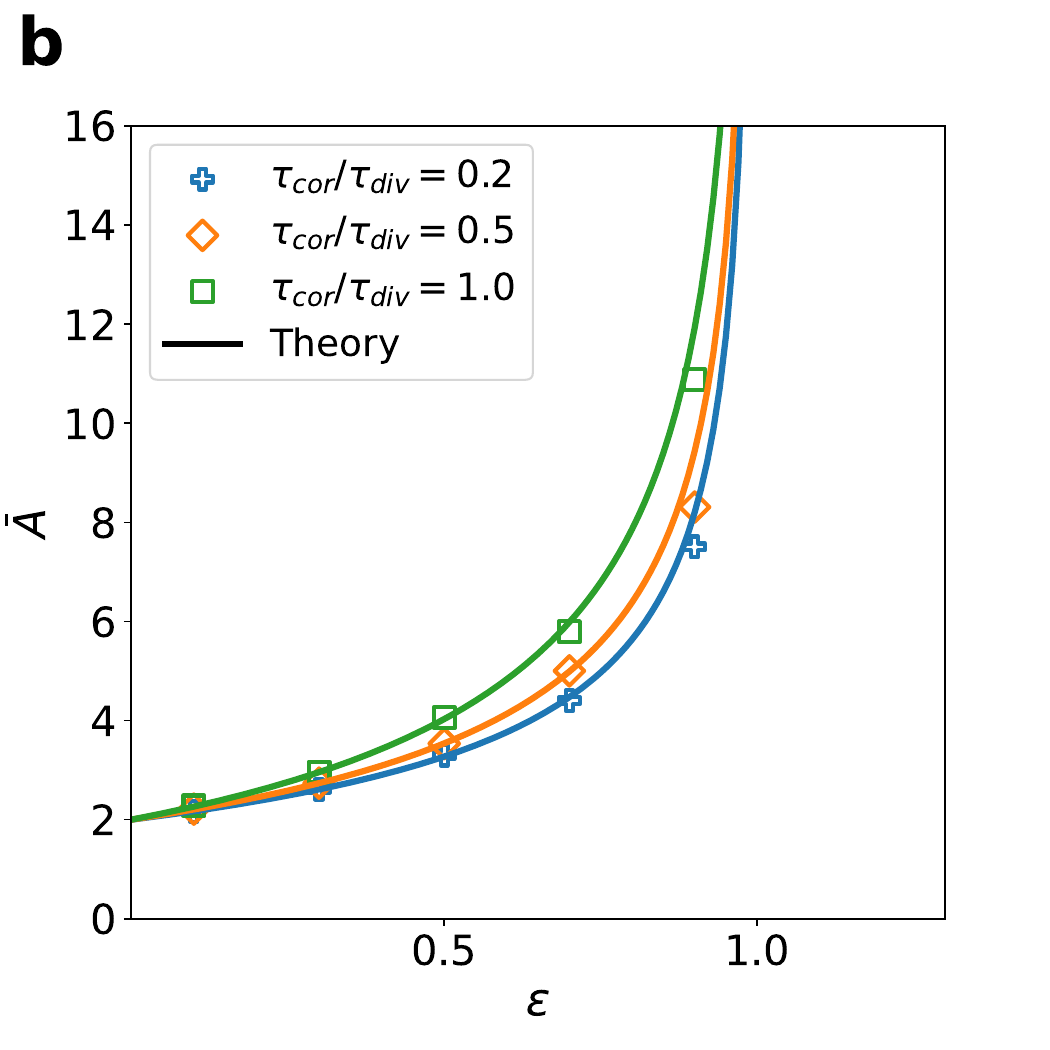}
\includegraphics[width=0.32\textwidth]{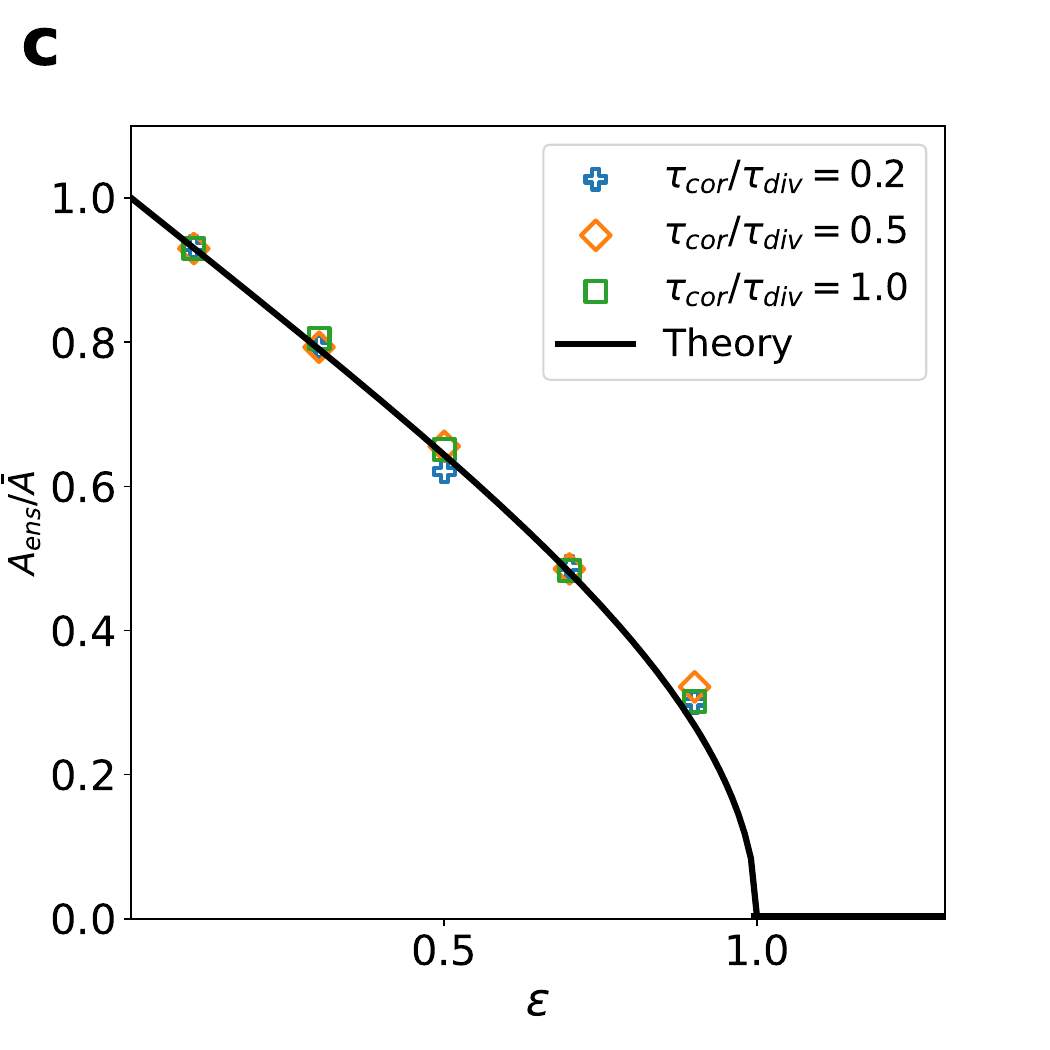}
\caption{Analytically calculated (solid curves) dependence of the typical amplitude $\bar A$, ensemble amplitude $A_{ens}$, and their ratio on the relative decay rate $\epsilon$ for various values of correlation times $\tau_{cor}$ compared to simulation results (markers). (a) The ensemble amplitude $A_{ens}$ has a relatively moderate dependence on $\epsilon$ and $\tau_{cor}$ (given in \eq{A_ens}). (b) The typical amplitude $\bar A$ has a strong dependence on $\epsilon$ (given in \eq{bar_A}) that diverges at $\epsilon=1$, and it decreases with $\tau_{cor}$. (c) When divided by each other, the $\epsilon$-dependence collapses on a single curve that is zero after the transition point $\epsilon=1$ and is given by \eq{order_param} before the transition. In that sense, the ratio $A_{ens}/\bar A$ is analogous to the order parameter in equilibrium phase transitions and is a measure of how deterministic the oscillations look. Note that numerically probing the vicinity of the transition point is more challenging as the required simulation time diverges at the transition point, and the number of cells grows exponentially with time. Simulation details: the pre-factors were estimated as $A\approx 2|\Psi(t)|e^{rt}$ (see \apx{osc_amp}), for large simulation time $t$. $\text{CV}_{m_d}$ is set to zero. For $\epsilon=0.1,0.3,0.5,0.7$, we used $1000$ independent simulations and a simulation time of $t=17\tau_{\text{div}}$. For $\epsilon=0.9$, we used $200$ independent simulations and a simulation time of $t=24\tau_{\text{div}}$.}
\label{fig:transition2}
\end{figure*}

We have seen in \eq{bar_A} that the typical amplitude $\bar A$ diverges as we approach $\epsilon=1$ (see also \fig{transition2} (b)). We can understand this by interpreting $A$ in \eq{Lt} as the Fourier amplitude $e^{rt}(\Lambda(t)-\Lambda_\infty)$ at frequency $\Omega$. As we approach the transition point, the contribution of the neglected fluctuations in \eq{Lt} to this Fourier mode increases and diverges after the transition point. This is because the neglected noise decays as $e^{-\Lambda_\infty/2}$ and when multiplied by $e^{rt}$ it becomes a growing function of time for $r>\Lambda_\infty/2$ (i.e. $\epsilon>1$). On the other hand, the ensemble amplitude $A_{ens}$ is a measure of this Fourier amplitude after the fluctuations in $N(t)$ are averaged out. With this in mind, we can interpret the ratio $A_{ens}/\bar A$ as a measure of how much of this Fourier mode is dominated by the deterministic oscillations. It is a measure of the order in this Fourier mode, sharply decays to zero at $\epsilon=1$, and stays zero after the transition point. It plays the same role that the order parameter plays in equilibrium phase transitions.

\Fig{transition2} shows the dependence of $A_{ens}$ and $\bar A$ and their ratio on $\epsilon$ and $\tau_{cor}$ (theory and simulation). Strikingly, the dependence of the \lq\lq order parameter\rq\rq $A_{ens}/\bar A$ on $\epsilon$ collapses for various model parameters into the same curve. The dependence of $A_{ens}$ and $\bar A$ on all other model parameters beside $\epsilon$ cancel out when dividing \eq{A_ens} by \eq{bar_A} giving the simple universal dependence of the order parameter on $\epsilon$ given by
\begin{align}\label{eq:order_param}
    \frac{A_{ens}}{\bar A} = \sqrt{2^{1-\epsilon}-1}.
\end{align}
This universal shape of the \lq\lq order parameter\rq\rq independent of model parameter exhibiting a sharp transition is analogous to the equilibrium phase transition. Similar to the equilibrium case, this sharp transition can only be observed in simulations at infinite system size (which corresponds to the large time limit in our growing system). Here, we have a nonequilibrium dynamics with a growing system size, for which (to the best of our knowledge) no theory of phase transition exists. All the comparisons with the equilibrium phase transitions are simple analogies. Some readers may prefer the language of abrupt nonlinear response to describe this behavior.

\section{Oscillation amplitude in a population starting from multiple cells}
For a general population starting from $N(0)$ cells with cell masses $m_1,\dots m_{N(0)}$, the oscillation amplitude pre-factor needs a slight modification
\begin{equation}
\label{eq:A_ens1}
    A_{\text{ens}} = 2e^{-41\text{CV}_{m_b}^2 + 123\sigma_\lambda^2\tau_{\text{cor}}^2} |\Psi(0)|.
\end{equation}
Here $\Psi(t)$ is the population's cell mass synchronicity defined as
\begin{equation}
\label{eq:Psi_def}
    \Psi(t) := \frac{1}{M(t)} \sum_{j=1}^{N(t)} m^{1+i\frac{2\pi}{\ln(2)}}_j, \quad M(t):= \sum_{j=1}^{N(t)}m_j(t).
\end{equation}
The absolute value $|\Psi(t)|$ ranges between $0$ and $1$ and measures the degree to which cell masses are synchronized. When $|\Psi(t)|=1$, all cells must have the same mass. In the special case of one cell, this always holds, hence \eq{A_ens1} reduces to \eq{A_ens}. When all cells are out of phase, we get $\Psi(t)=0$. In this case, there is full destructive interference in the first-order ensemble average oscillations of the future trajectory of $\Lambda_{\text{ens}}(t)$. Naturally, we have $\Psi(t)=0$ when cell masses are distributed according to the steady-state population mass distribution~\cite{hein2022asymptotic}. Whenever we talk about the instantaneous amplitude, we actually mean a re-scaled $|\Psi(t)|$. The general form of the typical amplitude is
\begin{equation}
    \bar A = 2e^{-41\text{CV}_{m_b}^2 + 123\sigma_\lambda^2\tau_{\text{cor}}^2} \sqrt{|\Psi(0)|^2 + J(0)}.
\end{equation}
Here $J(0)$ quantifies the net increase in transient oscillation amplitude with respect to the ensemble average due to early population stochasticity. A full formula is derived in the appendix, but it is closely approximated by
\begin{equation}
\label{eq:J0}
    J(0) \approx \frac{1}{M(0)} \left(\frac{1}{2^{1-\epsilon}-1}-1\right).
\end{equation}
 The larger the initial population, the smaller $J(0)$ and the smaller the effect of early population stochasticity. The strength of this effect also monotonically increases with growth rate variability. Interestingly, $J(0)$ is always a positive finite value, so $\bar A$ is strictly positive, even when the initial population is completely out of phase. This is demonstrated in \fig{A_dist} (c) and (d), where we provide the trajectories of instantaneous amplitudes as well as their distributions for a population starting from three out-of-phase cells. Note how in \fig{A_dist} (c), the small population immediately spontaneously synchronizes its cell masses, which then leads to sustained oscillations in the transient regime. When there is little to no synchronization in the initial population, we find that the asymptotic amplitude pre-factor $A$ of the later population follows a $\chi_2$ distribution, as discussed in Appendix~\ref{sec:multiple}. In \fig{A_dist} (d) we see that this distribution matches the simulated oscillation pre-factors very well. The instantaneous amplitude is defined as $A_{inst}(t)=e^{-41 \text{CV}_{m_d}}|\Psi(t)|$ and it is the best predictor for the amplitude of first order oscillations (see Appendix~\ref{sec:osc_fit} for a derivation and discussion).

\section{How long do the oscillations last?}
In this section, we consider how long it takes for the oscillation amplitude to fall below a certain threshold $A_{cut}$. The typical number of generations $g_{cut}$ to reach this threshold roughly satisfies $\bar A \exp[-g_{cut} \tau_{\text{div}}]=A_{cut}$. We can rewrite this as
\begin{equation}
   g_{cut} =  \frac{\bar\lambda}{r \ln(2)}\left[\ln(\bar A) - \ln(A_{cut})\right].
\end{equation}
Let us estimate what this amounts to for some experimentally obtained parameters, given a population that starts from a single cell with mass $\bar m$ at $t=0$. In \rf{taheri2015cell}, single lineage data is collected for E. \textit{coli} in various growth media. We are interested in the mean and coefficient of variation of the elongation rate and cell length at division, as well as the elongation rate's mother-daughter correlations. We assume cell length is proportional to cell mass and that the reported elongation rates are effective cell-cycle averages of some underlying Ornstein-Uhlenbeck growth rate process. Using theory from \rf{hein2022asymptotic}, we can deduce the set of parameters $\bar\lambda$, $\sigma_\lambda$, and $\tau_{\text{cor}}$ of the supposed underlying growth rate process, see \apx{params} for the relationships used. In Table \ref{tab:tab0} we report these values, derived from the experiments obtained over the seven different growth media. We also calculate the oscillation decay rate $r$, relative decay rate $\epsilon$, and typical amplitude $\bar A$. At last, we report the typical number of generations a population needs for the relative amplitude to fall below a cutoff of $0.1$. Note the large discrepancy ranging between $12$ and $34$ generations in this cutoff. This corresponds to a million-fold increase in population number. The growth medium thus has a large effect on whether the oscillations are observable or not.

\begin{table*}
    \begin{tabular}{ l | c | c|  c | c | c | c | c |}
      & TSB & Synth. rich & gcl+12a.a. & glc+6a.a & glucose & sorbitol & glycerol\\ \hline\\[-1.5\medskipamount]
      mean instantaneous single-cell growth rate, $\bar\lambda$ & 0.057 & 0.043 & 0.037 & 0.033 & 0.027 & 0.018 & 0.019 \\
      instantaneous growth rate variability, $\sigma_\lambda$ & 0.0060 & 0.0046 & 0.0035 & 0.0033 & 0.0027 & 0.0032 & 0.0024 \\
      correlation time of growth process, $\tau_{\text{cor}}$ & 6.1 & 6.1 & 4.4 & 7.0 & 5.9 & 9.6 & 18.8\\
      coefficient of variation of cell sizes at division, $\text{CV}_{m_d}$ & 0.17 & 0.14 & 0.10 & 0.10 & 0.10 & 0.12 & 0.11 \\ \hline 
      oscillation decay rate, $r$ & 0.018 & 0.010 & 0.0048 & 0.0061 & 0.0035 & 0.0081 & 0.0087 \\
      relative decay rate (oscillation/noise), $\epsilon$ & 0.63 & 0.48 & 0.24 & 0.37 & 0.26 & 0.86 & 0.91 \\
      typical oscillation amplitude, $\bar A$ & 1.4 & 1.5 & 1.6 & 1.9 & 1.7 & 3.8 & 6.3 \\
      number of generations until amplitude below a & 12 & 16 & 34 & 23 & 32 & 12 & 13\\
      \qquad cut-off, $g_{\text{cut}}$ for $A_{cut}=0.1$
    \end{tabular}
   \caption{Experimentally measured values of the model parameters for various growth conditions for \textit{E. coli} obtained from \rf{taheri2015cell} (see \apx{params} for details) together with analytically calculated values for oscillation decay rate $r$, relative decay rate $\epsilon$, typical amplitude pre-factor $\bar A$, and number of generation before the amplitude falls below $10\%$ of average growth rate. All seven growth conditions are below the transition points with "glucose + 12 amino acids" having the longest-lasting oscillations with amplitude larger than $10\%$, lasting for over 34 generations. See \rf{taheri2015cell} for the details of the growth conditions.}
    \label{tab:tab0}
\end{table*}

\section{Discussions and future directions}
In the 1950s, well before the existence of single-cell technology, microbiologists resorted to developing experimental methods to synchronize populations of growing bacteria to be able to probe single cells~\cite {lark1954induction,jun2018fundamental}. A celebrated example is what was widely known as the “baby machine” that screened cells of the same size to produce a synchronized population~\cite{helmstetter1963bacterial}. This method was famously used to discover multi-fork DNA replication in \textit{E. coli}~\cite{cooper1968chromosome}. These experiments inspired many early attempts to theoretically model the dynamics of synchronized growing populations~\cite{hirsch1966decay,bronk1968stochastic,brown1940note,kendall1948role}. These attempts use McKendrick-von Foerster formalism to analyze models akin to Bellman-Harris~\cite{bellman1952age}, where cells grow for a period of time independently drawn from a division time distribution and then divide. Such models are sometimes referred to as timer models or independent generation time models. These models predict that variation in generation times of cells (about $20\%$ variability~\cite{taheri2015cell}) desynchronizes the population (note that $20\%$ is well above the $13\%$ threshold in \eq{crossover_0}, so no one would have expected to get a synchronized population just by starting it from a single cell).

With the recent invention of single-cell Microfluidics technology such as the "mother machine"~\cite{wang2010robust} and subsequent theoretical developments in cell size regulation~\cite{amir2014cell}, we now understand that the aforementioned theoretical models fail to capture the oscillation dynamics quantitatively. This is because cell-size regulation drastically slows down the desynchronization~\cite{jafarpour2019cell}. There are two major sources of variability in the generation time of a cell: the variability in the growth rate, and the variability in the size at which the cells divide. The decay rate of the oscillations is set by the smaller source of noise, the noise in the growth rate of the cells. This is while, any long-term effect of the larger source of noise, the noise in the division size, on the population dynamics, is canceled by cell-size regulation. This comes with the realization that under conditions where the growth rate variability is sufficiently low, these oscillations can last longer than 30 generations. This is the time it takes for a $1$~ml culture, starting with a single cell, to saturate. In other words, all one needs to synchronize a large population is to start it from a single cell. This provides a way to probe hard-to-measure single-cell statistics by performing simple population measurements.

There have been very few experimental works on measurements of oscillations in population starting from single cells, and they only capture the very first few generations on gel plates using simple microscopy~\cite{chiorino2001desynchronization,vittadello2019mathematical}. Our theoretical predictions suggest that for many growth conditions, these oscillations can be very easily measured in liquid culture at the late stage when the population is growing deterministically. For this reason, we have written this paper with our experimental colleagues in mind to provide the necessary tools to help decide when these oscillations can be observed. 

It is important to realize that these oscillations are hidden from typical optical density measurements which estimate the total mass and not directly the cell count. The period of these oscillations is about one generation time, which means one needs multiple repeated measurements of instantaneous population growth rate per cell cycle (which could be as short as 23 min for \textit{E. coli}). The issue of time resolution can be alleviated by performing the experiment at a lower temperature, where all the processes in cells slow down almost proportionally. The population growth rate can be estimated by cell counting measurements performed twice on small samples a few minutes apart. We expect any well-designed experiment to be able to estimate the instantaneous population growth rate with uncertainty well below $10\%$. For this reason, we have estimated in Table \ref{tab:tab0} the number of generations it takes before the amplitude of the oscillations falls below $10\%$ of the mean growth rate. For certain growth conditions (glucose + 12 amino acids, see \rf{taheri2015cell} for the description of the growth media), the oscillations are predicted to maintain amplitude above $10\%$ of the mean growth rate for $34$ generations (this corresponds to longer than the time it takes a $10$~ml culture to saturate).

Here, we have shown that when the relative decay rate $\epsilon$ goes above 1, these oscillations are not observable. This threshold corresponds to roughly $13\%$ variability in growth rate, which is within the range $6-20\%$ variability reported in the literature~\cite{taheri2015cell,cermak2016high,kennard2016individuality,wallden2016synchronization}. However, careful estimations of the parameters of the model for all seven growth conditions from \rf{taheri2015cell} put \textit{E. coli} below this threshold (see Table \ref{tab:tab0}). We do not know if there is a biological significance for cells to stay below this transition point, and we have no reason to believe this is always the case. Moreover, due to the finite uncertainty in any measurement of cell counts, we find it unlikely for these oscillations to be observed in single experiments when $\epsilon$ is close to one which is the case for some of the growth conditions. 

In this work, we have tried to use a realistic model that keeps track of noise in both growth and division, is compatible with models of cell size regulation, and includes correlations in growth rates. One can always write more complicated models of growth and division keeping track of more degrees of freedom, but complicated models come with more parameters. We believe our model captures all the dominant sources of noise and correlations without involving parameters that are unknown or hard to measure. One source of noise and correlation that is being ignored in our model is any effect of asymmetry in cell division. \textit{E. coli} is extremely good at dividing symmetrically; the asymmetry in cell division of \textit{E. coli} is measured to have the relative variability of order $2\%$ under certain conditions~\cite{guberman2008psicic}, which would certainly have a negligible effect on any calculations performed in this paper. However, one could expect growth conditions or other species of bacteria in which there is a significant source of noise in the position of the division plane. The asymmetry in cell division could also manifest itself in an asymmetry in the division of the proteins in ribosomes among the two daughter cells which would lead to discontinuity in the growth process and subtle correlations between the growth rate and cell size~\cite{harsh2021non,nordholt2020biphasic}. Quantifying the effect of such asymmetries would be the subject of future work.

On the theoretical side, the competition between the transient dynamics and early stochasticity is what we believe to be a much more general phenomenon in stochastic exponential growth models. In most models of stochastic exponential growth, the early phases of growth are highly stochastic, but as the size of the system grows, the dynamics becomes more deterministic. In such cases, the effect of early stochasticity freezes in the distribution of the exponentially growing state. A well-known example of such dynamics is the Yule process, where the agents can spontaneously divide into two at a constant rate in a Markovian fashion. While the early stages of the Yule process are highly stochastic, it asymptotically leads to deterministic exponential growth with a pre-factor that approaches a stationary distribution. This pre-factor is the aforementioned frozen effect of stochasticity in the final state of the system. Similarly, Branching processes such as the Bellman-Harris model of population growth~\cite{bellman1952age} where the generation times are independently drawn from a fixed distribution, show early stochasticity that freezes in the final state of the system. Stochastic exponential growth can be also studied in the context of stochastic differential equations (SDEs) such as the Feller square root process~\cite{feller1951two} and its generalizations to power-law noise~\cite{cox1996constant}. Similar effects can be observed in these models as long as the stochasticity grows slower than the system size and they can be mapped to square root noise through a change of variable~\cite{pirjol2017phenomenology}. The exception is the case of geometric Brownian motion, where the noise grows proportional to the system's size~\cite{van1992stochastic}. Stochastic exponential growth is also observed in autocatalytic reaction networks such as those used in models of the origin of life~\cite{jafarpour2017noise,jafarpour2015noise} and models of single-cell growth~\cite{iyer2014universality,lin2020origin}. These models also have early stochasticity that grows asymptotically with the square root of the system size. 

While simpler models such as the Yule process and single-variable SDEs do not exhibit other transient dynamics besides the early stochasticity, the branching processes such as Bellman-Harris, and all the models with multiple degrees of freedom such as multivariate SDEs, autocatalytic networks, and models of single-cell growth can exhibit transient behavior which can be oscillatory. We believe the result of this work can and should be generalized to all such models where there is a transition between where the transient dynamics of macroscopic observable can be used to probe microscopic behavior and the regime where such microscopic information is masked by the early stochasticity. The model of growth and division in this paper is a complicated chimera of branching processes and SDEs. Nevertheless, we have solved this problem analytically and provided simple expressions for the transient dynamics in the presence of early stochasticity.

Besides in exponentially growing systems, both synchronized divisions and finite-size stochasticity are present in theoretical models and experimental setups of systems with finite population sizes such as the Moran process and turbidostat. It has recently been shown that the oscillations in the division rate in such systems persist indefinitely due to the finite coalescent time of finite populations~\cite{jafarpour2023evolutionary}. While the transition is not present in these systems due to the lack of exponential growth, we hope the techniques used in this paper prove fruitful in the analysis of such systems. Finite-size models are important in the context of population genetics since populations cannot grow exponentially over evolutionary timescales. Understanding the interaction of transient dynamics controlled by single-cell statistics and stochasticity in finite-size populations can provide insight into how evolution acts on cellular physiology. 

In many natural settings, bacteria grow in colonies with spatial structures. Within a spatially structured colony, cells that are closely related in their population tree are also closely related in space. This could create a strong local synchronization, despite the spatial heterogeneity that can desynchronize cells that are spatially separated. The interplay between this local synchrony and the mechanical and biological properties of the colony is a subject of future research.

\appendix

\section{The cell mass phase expansion}
Assume that the population is big enough to the point where the division and birth mass are distributed according to their steady-state distributions. Using Fourier expansion, we can rewrite the total cell count as an expansion in total cell mass and collective cell phase
\begin{align}
\label{eq:N_expansion}
    N(t) =& \frac{e^{\frac{1}{2}\text{CV}_{m_d}^2}}{2\ln(2) } \\ \nonumber
    \times &\left( M(t)+ \sum_{k=1}^\infty2e^{-\frac{1}{2}\omega^2 \text{CV}_{m_d}^2} \text{Re}\left[\frac{e^{ik\omega \text{CV}_{m_d}^2}}{1+ik\omega}\Phi_k(t) \right]\right),
\end{align}
where $\omega=2\pi/\ln(2)$ and total cell phase are given by
\begin{equation}
    \Phi_k (t) = \sum_{j=1}^{N(t)} m_{j}^{1+i\omega k}, \qquad M(t) = \Phi_0(t).
\end{equation}
In the next section, we will show that at large times, the total cell mass $M(t)$ grows exponentially at some rate $\Lambda_\infty$, and the collective mass phase terms $\Phi_k(t)$ grow exponentially while rotating in the complex plane. The higher order terms $\Phi_k(t)$ for $k\geq 2$ quickly become irrelevant, so we will neglect them. We will just consider the total cell mass $M(t)$ and first-order collective mass phase $\Phi(t)$, for which we drop the index.

\section{The transient dynamics of total mass and synchronicity}
In this section, we show that asymptotically, $\Phi(t)$ grows exponentially while rotating in the complex plane and show how to calculate these rates. We first define the collective mass-phase density of cells with growth state $\mathbf x$ as
\begin{equation}
    \Phi(t,\mathbf x) := \sum_{j=1}^{N(t)} m_j^{1+i\omega} \delta(\mathbf x_j -\mathbf x).
\end{equation}
Note that this quantity is conserved upon cell divisions.
The expected change thus purely depends on the change in cell mass $m_j$ and growth states $\mathbf x_j$ of individual cells
\begin{align}
     \left\langle \frac{d\Phi(t,\mathbf x) }{dt} \bigg |\Phi \right\rangle = & \sum_{j=1}^N \delta(\mathbf x_j-\mathbf x) \frac{d}{dt}\left\langle m^{1+i\omega} \big | m,\mathbf x\right\rangle  \notag \\
    & + \sum_{j=1}^{N(t)}  m_j^{1+i\omega} \frac{d}{dt}\left\langle\delta(\mathbf x_j-\mathbf x)|m,\mathbf x\right\rangle.
\end{align}
For the first term, we simply find $\partial_t m^{1+i\omega} = (1+i\omega)\lambda(\mathbf x) m^{1+i\omega}$, since the cell's state is conditioned to be at $\mathbf x$. The second term directly corresponds to the change in growth sate probability density $\partial_t p(t,\mathbf x)$, and we can thus use \eqref{eq:ddtp}. By combining the terms we get
\begin{equation}
\label{eq:dphidt}
    \left\langle \frac{d\Phi (t,\mathbf x)}{dt} \bigg |\Phi \right\rangle = \left(\mathcal{K} + (1+i\omega) \lambda(\mathbf x))\right) \Phi(t,\mathbf x).
\end{equation}
The steady-state dynamics of the collective phase are thus governed by $\langle \Phi(t,\mathbf x)\rangle \propto e^{\mu t}$, where $\mu$ is the leading complex-valued eigenvalue of the operator on the right-hand side of \eq{dphidt}. By noting that the total collective mass phase $\Phi(t)$ is simply the integral of $\Phi(t,\mathbf x)$ over all growth states, we find that its leading expected behavior must also evolve as $\langle \Phi(t)\rangle \propto e^{\mu t}$. Analogously we find that the expected total mass evolves as $\langle M(t)\rangle \propto e^{\Lambda_{\infty} t}$ where $\Lambda_\infty$ is the leading eigenvalue of $\mathcal K + \lambda(\mathbf x)$. By applying these transient results to the expectation of the first order expansion in \eq{N_expansion} we find that for large $t$, $\langle N(t)\rangle$ evolves as
\begin{equation}
\label{eq:EN}
    \langle N(t) \rangle \propto e^{\Lambda_\infty t}\left(1 + A_{\text{N,ens}} e^{-rt} \cos(\Omega t + \phi_{\text{N,ens}})\right),
\end{equation}
for some constants $A_{\text{N,ens}}$ and $\phi_{\text{N,ens}}$, where $r = \Lambda_\infty - \text{Re}\mu$ and $\Omega = \text{Im} \mu$. Plugging this into the definition of $\Lambda_{\text{ens}}(t)$ from \eqref{eq:LN} yields the first-order behavior of \eq{LEN} after ignoring terms of order $O(e^{-2rt})$.

\subsection*{The small growth rate variability limit}
\label{sec:small}
Based on \eq{dphidt}, one can show that the full solution to the expectation of the collective mass phase can be written as
\begin{equation}
\label{eq:EMPsi}
    \langle \Phi(t) \rangle = \left\langle e^{(1+i\omega)\int_0^t \lambda_sds} \right\rangle \Phi(0).
\end{equation}
where $\langle .\rangle$ is essentially a path integral over growth state space that integrates $\lambda_t$ over all growth rate trajectories along a lineage. The leading complex-valued constant $\mu$ for which asymptotically $\langle\Phi(t)\rangle \propto e^{\mu t}$ can be obtained from this expression via
\begin{equation}
\label{eq:mu0}
    \mu =\lim_{t\to\infty} \frac{1}{t} \ln \left\langle e^{(1+i\omega)\int_0^t \lambda_s ds} \right\rangle.
\end{equation}
When the growth rate integral is Gaussian (which holds when $\lambda_t$ is Ornstein-Uhlenbeck) or variations in $\lambda_s$ are small, one can perform second-order cumulant expansion on \eq{mu0}. to obtain
\begin{equation}
\label{eq:mu1}
    \mu = (1+i\omega) \bar\lambda + (1+i\omega)^2 D.
\end{equation}
Here, the lineage steady-state growth rate is 
\begin{equation}
    \bar\lambda := \lim_{t\to \infty} \frac{1}{t} \left\langle \int_0^t \lambda_sds\right\rangle = \langle\lambda_0\rangle_{ss},
\end{equation}
where the subscript $ss$ denotes that $\lambda_0$ is taken from a lineage steady-state distribution. The growth accumulation diffusion constant $D$ is
\begin{equation}
D:=\lim_{t\to\infty} \frac{1}{2t} \text{Var}\left(\int_0^t \lambda_s ds\right) = \int_0^\infty \text{Cov}\left(\lambda_0,\lambda_s\right)_{ss} ds.
\end{equation}
In an analogous derivation where we substitute $\omega\to 0$, one obtains the asymptotic population growth rate $\Lambda_\infty = \bar\lambda + D$. The population time scales given in Section \ref{sec:Ens} can now easily be derived.

\section{The binned population growth rate}
\label{sec:envelopes}
In this section, we discuss how we estimated the population growth rate, as well as oscillation fits and fluctuation envelopes shown in \fig{transition} and \fig{osc_fits}. Naturally, when dealing with finite discrete populations, the population growth rate in \eq{Lt_def} is not well defined, since it contains the derivative of a discrete variable. We thus consider small but finite time intervals $\Delta t$. The discretized population growth rate shown in \fig{transition} and \fig{osc_fits} is defined as 
\begin{equation}
\label{eq:L_Delta_def}
    \Lambda_{\Delta t} = \frac{\ln(N(t+\Delta t/2)) -\ln(N(t-\Delta t/2))}{\Delta t}
\end{equation}

\subsection{The fluctuation envelope}
\label{sec:fluctuation envelope}
In the limit of a population completely dominated by random fluctuations, all collective synchronicity disappears and we may assume the cells in the population to be independent. Consider a population of $N(t-\Delta t/2)$ independent cells. Given a small enough time interval $\Delta t$, each cell randomly chooses to divide or not divide with the same probability $p$. The number of cells after this time interval $N(t+\Delta t/2)$ can now be related to the previous population via
\begin{equation}
    N(t+\Delta t/2) = N(t-\Delta t/2) + \mathcal{B}
\end{equation}
where $\mathcal{B} \sim \text{Binomial}(N(t-\Delta t/2),p)$. Assuming that the expected population predictably grows as $\langle N(t+\Delta t/2)\rangle = e^{\Lambda\infty \Delta t}\langle N(t-\Delta t/2)\rangle$, we can derive the probability that an individual cell divides as $p\approx \Lambda_\infty\Delta t $. The discretized population growth rate can be expressed as
\begin{equation}
\label{eq:L_Delta}
    \Lambda_{\Delta t} = \frac{1}{\Delta t} \ln\left( 1+ \frac{\mathcal{B}}{N(t-\Delta t/2)}\right)
\end{equation}
By using properties of the binomial distribution, one can show that up to leading order in small $\Delta t$, typical fluctuations in $\Lambda_{\Delta t}$ scale with
\begin{equation}
    \sqrt{ \text{Var} \left(\Lambda_{\Delta t} \right)} \approx \sqrt{ \frac{\Lambda_\infty}{N(t) \Delta t}}.
\end{equation}
To smoothen out this fluctuation envelope, we approximated the cell count by the scaled total cell mass $N(t) \approx M(t)/2\ln(2)$, to obtain
\begin{equation}
    \sqrt{\text{Var}\left(\Lambda_{\Delta t}\right)}\approx \sqrt{\frac{\Lambda_\infty 2\ln(2)}{M(t) \Delta t}}.
\end{equation}
This is what we used to determine the fluctuation envelope in \fig{transition} (c) and (d).

\subsection{The oscillation fit and envelope}
\label{sec:osc_fit}
In this section, we explain how we determined the oscillation fits and envelopes \fig{transition} and \fig{osc_fits}. Usually, a fit would involve a procedure that minimizes the distance between data and a parametrized curve. Our method instead employs the idea that oscillation phase and amplitude can be directly deduced from the population cell mass distribution at any point in time. Since the oscillations become more and more deterministic as the population gets larger, the best estimate of the transient limit amplitude and phase are based on the mass distribution at the largest simulated time $t_{max}$ available. 

In the limit of a large population, the total cell mass and collective mass phase will approach the dynamics of their expectation values, hence for large $t$ we have $\partial_t M(t) = \Lambda_\infty M(t)$ and $\partial_t \Phi(t) =\mu \Phi(t)$. For all large $t$ we can reasonably extrapolate these deterministic dynamics around the population's state at some large time $t_{max}$ to obtain 
\begin{equation}
    M(t) \approx M(t_{max})e^{\Lambda_\infty(t-t_{max})},
\end{equation}
and
\begin{equation}
     \Phi(t) \approx \Phi(t_{max})e^{\mu(t-t_{max})}.
\end{equation}
Subsequently, we can plug these approximations into \eq{L_Delta_def} using a first-order expansion of \eq{N_expansion} to find
\begin{equation}
\label{eq:L_Delta_sol}
    \Lambda_{\Delta t}\approx \Lambda_\infty \left( 1+ \text{Re}\left[ C_{\Delta t} \Psi(t_{max}) e^{(-r+i\Omega)(t-t_{max})}\right]\right),
\end{equation}
where $\Psi(t)=\Phi(t)/M(t)$ and 
\begin{equation}
\label{eq:C_def}
    C_{\Delta t} = 2e^{(i\omega -\frac{1}{2}\omega^2)\text{CV}_{m_d}^2} \frac{e^{(-r+i\Omega)\Delta t/2}-e^{-(-r+i\Omega)\Delta t/2}}{\Delta t(1+i\omega)\Lambda_\infty},
\end{equation}
in the main text, we use the approximation 
\begin{equation}
\label{eq:C_approx}
    C_{\Delta t} \approx 2e^{(i\omega -\frac{1}{2}\omega^2)\text{CV}_{m_d}^2},
\end{equation}
which is valid for small $D$ and $\Delta t$. Determining the envelope of the oscillations in \eq{L_Delta_sol} is straightforward: You take the absolute value of the term that rotates in the complex plane. The upper envelope is thus given by. 
\begin{equation}
\label{eq:L_env}
    \Lambda_{env} \approx \Lambda_\infty\left(1+ |C_{\Delta t} \Psi(t_{max})| e^{-r(t-t_{max})}\right)
\end{equation}
Subsequently, we define the instantaneous amplitude at time $t$ as 
\begin{equation}
    A_{\text{inst}}(t) = |C_{\Delta t}\Psi(t)|
\end{equation}
Given a population at time $t_{max}$ with dynamics that are like a large population, the best estimate for the amplitude of future oscillations is $A_{\text{inst}}(t_{max})e^{-r(t-t_{max})}$.

\section{The amplitude distribution}
\label{sec:osc_amp}
In the previous section, we discussed how we can fit transient oscillations to the population growth rate based on the instantaneous complex amplitude $C_{\Delta t}\Psi(t_{max})$ at some large time $t_{max}$. As we increase $t_{max}$, \eq{L_Delta_sol} becomes an increasingly better descriptor of the transient oscillations at large $t$. This is demonstrated in \fig{A_dist} (a) and (c), where we show trajectories of the instantaneous amplitudes $A_{\text{inst}}(t)=|C_{\Delta t}\Psi(t)|$ of independent populations at finite times. Importantly, we see that at large $t$, all trajectories approach a curve of the form $Ae^{-rt}$. Similarly, the complex instantaneous amplitude $C_{\Delta t}\Psi(t)$ approaches a complex curve of the form $Ae^{i\phi} e^{-(r-i\Omega)t}$. Consequently, \eq{L_Delta_sol} can be written as
\begin{equation}
    \Lambda(t) \approx \left(1+ Ae^{-r t}\cos(\Omega t+\phi)\right),
\end{equation}
where amplitude pre-factor $A$ and phase $\phi$ can be expressed as the limit
\begin{equation}
\label{eq:Aphi_lim}
    A e^{i\phi} = \lim_{t\to\infty} C_{\Delta t}\Psi(t) e^{(r-i\Omega)t}.
\end{equation}
In this section, we will use this limit to determine the typical transient amplitude $\bar A=\sqrt{\langle A^2\rangle}$ and its distribution in a special case where initial cell masses are out of phase. In \eq{Aphi_lim} we see that $A e^{i\phi}$ is proportional to $\Psi(t)=\Phi(t)/M(t)$ at large times $t$, where we recall that $\Phi(t)=\sum_j m_j^{1+i\omega}$ is the collective mass phase and $M(t)=\sum_j m_j$ is the total cell mass. Variability in $M(t)$ is very small compared to the variability in $\Phi(t)$, so we will approximate the total cell mass by its expected value $\Psi(t) \approx \Phi(t)/\langle M(t)\rangle$. Our analysis of $A$ will thus use the expression
\begin{equation}
\label{eq:Aphi_lim_approx}
    Ae^{i\phi} \approx \lim_{t\to\infty} C_{\Delta t} \frac{\Phi(t)}{\langle M(t)\rangle} e^{(r-i\Omega)t}
\end{equation}
$A$ and $\phi$ are closely related to the amplitude $A_{\text{ens}}$ and phase $\phi_{\text{ens}}$ of the ensemble population.

By taking the expectation value on both sides of \eq{N_expansion} and plugging the first-order expansion into \eq{LN}, we obtain first-order oscillatory behavior of the form \eq{LEN}, where the ensemble amplitude $A_{\text{ens}}$ and phase $\phi_{\text{ens}}$ are given by
\begin{equation}
\label{eq:A_ens_lim}
    A_{\text{ens}}e^{i\phi_{\text{ens}}} = \lim_{t\to\infty}  C_{\Delta t}  \frac{\langle \Phi(t)\rangle}{\langle M(t)\rangle} e^{(r-i\Omega)t}
\end{equation}
This expression can be solved exactly when $\lambda_t$ is an Ornstein-Uhlenbeck process. In this case, we know that $\int_0^t\lambda_sds$ is normally distributed, where for large $t$ the mean and variance are
\begin{equation}
    \left\langle \int_0^t\lambda_sds\right\rangle = \bar\lambda t + (\lambda_0-\bar\lambda)\tau_{\text{cor}},
\end{equation}
and 
\begin{equation}
\label{eq:var}
    \text{Var}\left( \int_0^t\lambda_sds\right)= 2D t - 3D\tau_{\text{cor}}.
\end{equation}
note that both of these are proportional to $t$, with some shift that scales with the time $\tau_{\text{cor}}$ it takes for the distribution of $\lambda_t$ to go to its steady state. We can now calculate the expected values of total mass and collective mass phase using \eq{EMPsi} and obtain the ensemble amplitude from \eq{A_ens_lim},
\begin{equation}
    A_{\text{ens}} = e^{ \frac{3}{2}\omega^2 \sigma_\lambda^2\tau_{\text{cor}}^2} |C_{\Delta t}\Psi(0)|.
\end{equation}
The term $\frac{3}{2}\omega^2 \sigma_\lambda^2\tau_{\text{cor}}^2$ results from the constant shift in \eq{var}. From \eq{Aphi_lim_approx} and \eq{A_ens_lim} we can tell that $\langle A e^{i\phi} \rangle \approx A_{\text{ens}}e^{i\phi_{\text{ens}}}$. This already gives us the first indication that the average amplitude of a single simulation is strictly higher than the ensemble average amplitude, by noting that
\begin{equation}
    \langle A\rangle = \langle |A e^{i\phi}|\rangle > |\langle A e^{i\phi} \rangle | \approx A_{\text{ens}}.
\end{equation}
The higher the spread in $\phi$, the bigger the difference between $\langle A\rangle$ and $A_{\text{ens}}$.
Although we cannot calculate the average amplitude $\langle A\rangle$ directly, we can use some tricks to find its average square $\langle A^2\rangle$ instead. The ratio between the typical amplitude and ensemble amplitude $\sqrt{\langle A^2\rangle}/A_{ens}$ has a negligible dependence on growth and division variability, so we will consider 
a special limit in which the ratio can be calculated analytically and apply that result generally. We will assume $\bar\lambda \tau_{\text{corr}}\ll 1$, with finite $D=\sigma_\lambda^2 \tau_{\text{cor}}$ , and that cells always divide upon attaining a cell mass of $2$. In the limit of small growth rate correlations, the integrated growth rate $\int_0^t\lambda_s ds$ will be identical to a Brownian motion with drift, with diffusion constant $D$ and drift velocity $\bar\lambda$. For a cell with initial mass $m_0$, its mass at time $t$ is given by
\begin{equation}
    m(t) = m_0 e^{\int_0^t \lambda_sds}.
\end{equation}
The time for a cell of initial mass $m_0$ to reach $2$ and divide is $T_{m_0}=T(\ln(2/m_0))$ where
\begin{equation}
    T(u) := \left\{ t>0 : \int_0^t \lambda_sds \geq u\right\}.
\end{equation}
This is equivalent to the hitting time of a Brownian motion with drift, which is known to have an inverse Gaussian distribution. This gives us a moment-generating function of the division time
\begin{equation}
\label{eq:generating}
    \left\langle e^{aT(u)}\right\rangle = \exp\left[\frac{u}{2 D}\left(\bar\lambda - \sqrt{\bar\lambda^2-4D a}\right)\right].
\end{equation}

Consider the asymptotic collective mass phase prefactor of a population starting from one cell with mass $m_0$
\begin{equation}
\label{eq:Z_def}
    Z(m_0):= \lim_{t\to\infty} \frac{e^{-\mu t} \Phi(m_0,t)}{m_0^{1+i\omega}},
\end{equation}
which is a complex-valued random variable with $\langle Z(m_0)\rangle=1$. We use notation $Z=Z(1)$.
The first division takes place at $t=T_{m_0}$, at which point the population splits up into two identical sub-populations that start with mass $1$ at time $t=T_{m_0}$. This gives us a relationship for the collective mass phase similar to the one used to derive Powell's relationship \cite{powell1956growth}. 
\begin{equation}
    \Phi(m_0,t) = \Phi_\uparrow\left(t-T_{m_0}\right) + \Phi_\downarrow\left(t-T_{m_0}\right),
\end{equation}
where the arrows are used to distinguish the two independent sub-populations. Plugging this relationship into \eq{Z_def} we find
\begin{equation}
    Z(m_0) := \frac{1}{m_0^{1+i\omega}} \left( Z_\uparrow + Z_\downarrow\right) e^{-\mu T_{m_0}}
\end{equation}
Now we multiply both sides by their conjugate and take their expected values to find
\begin{equation}
\label{eq:Z_rec0}
    \left\langle |Z(m_0)|^2\right\rangle= \frac{1}{m_0^2} \left( 2\left\langle |Z|^2\right\rangle + 2 \right) \left\langle e^{-2\text{Re} \mu T_{m_0}}\right\rangle.
\end{equation}
With the help of \eq{generating} we can show that
\begin{equation}
    \left\langle e^{-2\text{Re} \mu T(u)}\right\rangle = e^{(-2+\epsilon^*) u},
\end{equation}
where
\begin{equation}
\label{eq:eps_prime}
    \epsilon^* := 2- \frac{1}{2D}\left( \sqrt{\bar\lambda^2 + 8D\bar\lambda + 8(1-\omega^2)D^2} - \bar\lambda^2\right).
\end{equation}
One can show that up to the first order in $D/\bar\lambda$ we have
\begin{equation}
    \epsilon^* \approx 2\omega^2 D/\bar\lambda \approx \epsilon
\end{equation}
This lets us rewrite \eq{Z_rec0} as 
\begin{equation}
\label{eq:Z_rec1}
    \left\langle |Z(m_0)|^2\right\rangle= \frac{1}{2} \left( \left\langle |Z|^2\right\rangle + 1 \right) e^{\epsilon^* (\ln(2)-\ln(m_0))}.
\end{equation}
If the initial population starts from one cell with mass $m_0=1$, then its value of $Z(1)$ will be equal in distribution to that of either of the daughter populations. This way one can recursively solve for $\left\langle |Z|^2\right\rangle$ in \eq{Z_rec1}. The full solution for arbitrary initial cell mass is
\begin{equation}
\label{eq:Z_sol}
    \left\langle |Z(m_0)|^2\right\rangle = \frac{m_0^{\epsilon^*}}{2^{1-\epsilon^*}-1}.
\end{equation}
This is the final ingredient needed to solve the value of the typical amplitude since we can write

\begin{equation}
\label{eq:A_ratio}
    \frac{\bar A}{A_{\text{ens}}} \approx \lim_{t\to\infty} \frac{\sqrt{\langle|\Phi(t)|^2\rangle}}{|\langle \Phi(t)\rangle|} = \sqrt{\langle |Z(m_0)|^2\rangle}.
\end{equation}
We now use use \eq{Z_sol} to find 
\begin{equation}
    \frac{\bar A}{A_{\text{ens} }}\approx \sqrt{\frac{m_0^{\epsilon}}{2^{1-\epsilon}-1}}.
\end{equation}
which gives the typical amplitude for a population stemming from one cell.
\subsection{Multiple cells in initial population}\label{sec:multiple}

When the population starts from multiple cells with masses $m_j$, the collective mass phase is a complex superposition of the mass phase trajectories of the populations stemming from each of the individual cells $\Phi_j(m_j,t)$, so
\begin{equation}
\Phi_{\text{total}}(t) = \sum_{j=1}^{N(0)}\Phi_j(m_j,t).
\end{equation}
Note that the different mass phase trajectories $\Phi_j(m_j,t)$ are independent. We can thus plug this into \eqref{eq:Aphi_lim_approx} and take the expected absolute square to find
\begin{equation}
 \bar A = |C_{\Delta t}|e^{\frac{3}{2}\omega^2\sigma_\lambda^2\tau_{\text{cor}}^2} \sqrt{|\Psi(0)|^2 + J(0)},
\end{equation}
where
\begin{equation}
\label{eq:J1}
    J(0)= \frac{1}{M(0)^2 } \sum_{j=1}^{N(0)} m_{j}^2 \left(\frac{m_j^{-\epsilon}}{2^{1-\epsilon}-1} -1\right).
\end{equation}
For all values of $\epsilon$ and $m_j$, this is well approximated by \eq{J0} in the main text. We also approximated $C_{\Delta t}$ by \eq{C_approx}. When the initial population has a well-mixed mass distribution and $\Phi_{\text{total}}(0)$ is close to zero, one could argue with the help of a central limit theorem that $\Phi_{\text{total}}(t)$ and $Ae^{i\phi}$ have a centered rotationally symmetric bi-variate Gaussian distribution in the complex plane. In that case, we know that $A$ itself must have a $\chi_2$-distribution, with a probability density function of 
\begin{equation}
    f_{A}(a) = \frac{2 a}{\bar A^2} e^{-\frac{a^2}{\bar A^2}}.
\end{equation}
The notion that this is the amplitude distribution for a well-mixed initial population is supported by \fig{A_dist} (d), where we see that it is a great fit for a population starting from as few as just three cells.

\begin{table*}[t]
\begin{tabular}{ c | c | c|  c | c | c | c | c |}
  & TSB & Synth. rich & gcl+12a.a. & glc+6a.a & glucose & sorbitol & glycerol\\ \hline
  $\bar\kappa$ & 0.057 & 0.043 & 0.037 & 0.033 & 0.027 & 0.018 & 0.019 \\
  $\sigma_\kappa$ & 0.0045 & 0.0032 & 0.0021 & 0.0022 & 0.0016 & 0.0020 & 0.0018 \\
  $\rho_{m-d}$ & 0.33 & 0.26 & 0.14 & 0.22 & 0.14 & 0.17 & 0.34\\
$\text{CV}_{m_d}$ & 0.17 & 0.14 & 0.10 & 0.10 & 0.10 & 0.12 & 0.11
\end{tabular}
\caption{Experimentally measured values of the mean, standard deviation, and mother-daughter correlations of per cell growth rates for E. coli in several growth media as reported in Table S3 of \rf{taheri2015cell}. The mother-daughter correlations were obtained from the elongation rates in Figure S4 of \rf{taheri2015cell} by taking the first-generation correlations and subtracting the average of the three final correlations as extrinsic noise.}
\label{tab:tab1}
\end{table*}

\section{Determining the growth process parameters}
\label{sec:params}
In \rf{hein2022asymptotic}, a relationship is derived between parameters of an Ornstein-Uhlenbeck growth process $\bar\lambda$, $\sigma_\lambda$, $\tau_{\text{cor}}$  and the emergent mean, $\bar\kappa$, standard-deviation $\sigma_\kappa$ and mother-daughter correlations $\rho_{m-d}$ of growth rate averaged over cell cycles. Up to first order in $\sigma_\lambda^2/\bar\lambda^2$, their results are
\begin{equation}
\label{eq:kappa_bar0}
    \bar \kappa = \bar\lambda\left(1+ \frac{\sigma_\lambda^2}{\bar\lambda} h\left(\frac{\ln(2)}{\bar\lambda \tau_{\text{cor}}}\right)\right),
\end{equation}

\begin{equation}
\label{eq:kappa_var0}
    \sigma_\kappa^2 = \sigma_\lambda^2 h\left(\frac{\ln(2)}{\bar\lambda \tau_{\text{cor}}}\right),
\end{equation}

\begin{equation}
\label{eq:kappa_cor0}
    \rho_{m-d}= h_2\left(\frac{\ln(2)}{\bar\lambda \tau_{\text{cor}}}\right),
\end{equation}
where $h_1(z)$ and $h_2(z)$ are auxiliary functions defined as
\begin{equation}
h_1(z) = \frac{2}{z}\left(1-\frac{1}{z}(1-e^{-z})\right),
\end{equation}
\begin{equation}
    h_2(z) = \frac{1}{2} \frac{(1-e^{-z})^2}{z-(1-e^{-z})}.
\end{equation}
By inverting this set of equations we obtained the Ornstein-Uhlenbeck process parameters in Table \ref{tab:tab0} based on values from \rf{taheri2015cell}. The values used are listed in Table \ref{tab:tab1}.

\subsection*{The transition point for vanishing growth rate correlations}
\label{sec:crossover_0}
To show that our results are consistent with the transition point derived in the introduction, we want to take a limit of the continuous growth rate model that corresponds to the per-cell growth rate model with vanishing correlations from \rf{jafarpour2019cell}. This can be achieved by taking $\bar\lambda \tau_{\text{cor}} \ll 1$ while fixing $D=\sigma_\lambda^2 \tau_{\text{cor}}/\bar\lambda$. In this limit we obtain 
\begin{equation}
\label{eq:kappa_mean1}
    \bar\kappa = \bar\lambda \left( 1+ \frac{2}{\ln(2)} \frac{D}{\bar\lambda}\right),
\end{equation}
and
\begin{equation}
\label{eq:kappa_var1}
    \sigma_\kappa^2 = \bar\lambda^2 \frac{2}{\ln(2)} \frac{D}{\bar\lambda},
\end{equation}
and $\rho_{m-d}\ll 1$. Recall that for a continuous process, the transition point was given by $164D\leq\bar\lambda$. By using the conversion given by \eq{kappa_mean1} and \eq{kappa_var1}, we recover the transition point given in \eq{crossover_0} in the introduction.
\begin{equation}
    \sigma_\kappa \leq 0.13\bar\kappa.
\end{equation}

\begin{acknowledgments}
We gratefully acknowledge Michael Vennettilli and Mahshid Jafarpour for their valuable feedback on the manuscript.
\end{acknowledgments}

\bibliography{ref}

\end{document}